\documentclass[aps,prd,preprintnumbers,nofootinbib,superscriptaddress]{revtex4}
\usepackage{amsmath}

\usepackage{epsfig}
\usepackage{amssymb,bm,bbm,color}
\usepackage{graphicx}
\usepackage{verbatim}
\usepackage{simplewick}
\usepackage{appendix}

\def\beq{\begin{equation}}
\def\eeq{\end{equation}}
\newcommand{\bea}{\begin{eqnarray}}
\newcommand{\eea}{\end{eqnarray}}
\def\bi{\begin{itemize}}
\def\ei{\end{itemize}}
\def\ba{\begin{array}}
\def\ea{\end{array}}
\def\bfig{\begin{figure}}
\def\efig{\end{figure}}
\def\p{\partial}
\def\d{\delta}
\def\U{{\cal U}}
\def\E{{\cal E}}
\def\K{{\cal K}}
\def\g{{\gamma}}

\newcommand{\calE}{{\cal E}}
\newcommand{\calK}{{\cal K}}

\newcommand{\calU}{{\cal U}}
\newcommand{\dd}{{\rm d}}

\newcommand{\tk}{\tilde k}

\newcommand{\f}{e}

\begin{document}

\preprint{IPMU14-0108}

\title{Cosmology in rotation-invariant massive gravity with non-trivial fiducial metric}
\author{David  Langlois} 
\affiliation{APC (CNRS-Universit\'e Paris 7), 10 rue Alice Domon et L\'eonie Duquet, 75205 Paris Cedex 13, France}

\author{Shinji Mukohyama}
\affiliation{Kavli Institute for the Physics and Mathematics of the Universe (WPI), Todai Institutes for Advanced Study, The University of Tokyo, 5-1-5 Kashiwanoha, Kashiwa, Chiba 277-8583, Japan}

\author{Ryo Namba}
\affiliation{Kavli Institute for the Physics and Mathematics of the Universe (WPI), Todai Institutes for Advanced Study, The University of Tokyo, 5-1-5 Kashiwanoha, Kashiwa, Chiba 277-8583, Japan}

\author{Atsushi Naruko}
\affiliation{APC (CNRS-Universit\'e Paris 7), 10 rue Alice Domon et L\'eonie Duquet, 75205 Paris Cedex 13, France}
\affiliation{Department of Physics, Tokyo Institute of Technology, Tokyo 152-8551, Japan}

\date{\today}

\begin{abstract}
We investigate the cosmology of $SO(3)$-invariant massive gravity with $5$ degrees of freedom. In contrast with previous studies, we  allow for  a non-trivial fiducial metric, which can be justified  by invoking, for example, a dilaton-like global symmetry. We write the homogeneous and isotropic equations of motion in this more general setup and identify, in particular, de Sitter solutions. We then study the linear perturbations around the homogeneous cosmological solutions, by deriving the quadratic Lagrangian governing the dynamics of scalar, vector and tensor modes. We thus obtain the conditions for the perturbations to be well-behaved.  We show that it is possible to find de Sitter solutions whose perturbations are weakly coupled and stable, i.e. without ghost-like or gradient instabilities. 
\end{abstract}

\maketitle

\section{Introduction}

Open questions  in modern cosmology such as the origin of the
accelerating universe or of the flattening of galaxy rotation curves
have provided strong motivation to study modifications of gravity in the
infrared (IR). While, in the usual explanation,  dark energy and dark
matter are responsible for those interesting phenomena, it is also
natural to ask whether one can change the behavior of gravity in the IR
from  general relativity (GR) to address these mysteries.

One possibility to modify gravity in the IR is to give a mass to the
graviton. Theoretically speaking, however, it has been a long standing
problem whether the graviton can have a non-vanishing mass. While Fierz
and Pauli (FP) in 1939~\cite{Fierz:1939ix} found the unique
Lorentz-invariant linear theory of massive gravity without ghost
instability in Minkowski background, it was found in 1970 by van Dam,
Veltman and Zakharov (vDVZ)~\cite{vanDam:1970vg,Zakharov:1970cc} that
the  massless limit of the FP theory does not reproduce the massless
theory, i.e. GR. Indeed, the post-Newtonian parameter $\gamma$ in the
massless limit is $1/2$ and does not agree with its GR value of
$1$. Hence, if the linear theory is valid in the massless limit then
massive gravity would have been experimentally excluded, however small
the graviton mass is. In 1972 Vainshtein~\cite{Vainshtein:1972sx} then
showed that the linear approximation breaks down in the massless limit
and thus the vDVZ discontinuity is not physical. He further argued that
because of nonlinearity the massless limit of massive gravity recovers
GR in the region where nonlinearity is significant enough. The radius of
this region for a given gravitational source is often called the
Vainshtein radius, and can be made arbitrarily large by making the
graviton mass sufficiently small. However, in the same year Boulware and 
Deser~\cite{Boulware:1973my} showed that nonlinear extensions of the FP
theory generically exhibit ghost instability. (See reviews
\cite{Hinterbichler:2011tt,deRham:2014zqa} for more details of the
history and various issues of massive gravity.) Since then, the theory
of massive gravity has been haunted by this ghost, often called the
Boulware-Deser (BD) ghost, for almost $40$ years until de Rham,
Gabadadze and Tolley (dRGT)~\cite{deRham:2010ik,deRham:2010kj} finally 
found, in 2010, a fully nonlinear theory of massive gravity without BD
ghost. While the dRGT theory was constructed in such a way that it
avoids the BD ghost in the so-called decoupling limit, the general proof
of the absence of BD ghost away from the decoupling limit was completed
by Hassan and Rosen~\cite{Hassan:2011hr,Hassan:2011ea} (see also 
\cite{Kugo:2014hja}).

Having a promising candidate for a theoretically consistent nonlinear theory
of massive gravity, it is natural to investigate its cosmological 
implications. However, it was soon uncovered that the original dRGT
theory does not admit any nontrivial flat homogeneous and isotropic
Friedmann-Lema\^\i tre-Robertson-Walker (FLRW)
solution~\cite{D'Amico:2011jj}. While this no-go result extends to the
closed FLRW case, self-accelerating open FLRW solutions were found in
\cite{Gumrukcuoglu:2011ew}. The negative spatial curvature of this
solution can be made arbitrarily close to zero, and  current
observational data are consistent with a negative curvature of a percent
level. Moreover, by replacing the Minkowski fiducial
metric~\footnote{The covariant formulation of the dRGT theory  
has a physical metric and four St\"{u}ckelberg scalar fields. The
pullback of the metric in the space of St\"{u}ckelberg scalar fields
into the physical spacetime is called the fiducial metric.} in the dRGT
theory with a de Sitter or FLRW fiducial metric, one can find both
self-accelerating~\cite{Gumrukcuoglu:2011zh} and non 
self-accelerating~\cite{Hassan:2011vm,Fasiello:2012rw,Langlois:2012hk} 
solutions of not only open but also flat and closed FLRW
types. Unfortunately, it was later found in \cite{DeFelice:2012mx} that
all homogeneous and isotropic FLRW solutions in dRGT theory are unstable
due to either a linear ghost called Higuchi ghost~\cite{Higuchi:1986py} or
a new type of nonlinear ghost. The new nonlinear ghost instability stems
from the fact that three among five physical degrees of freedom have
vanishing (time) kinetic terms on the self-accelerating FLRW
solutions~\cite{Gumrukcuoglu:2011zh}.

In principle this no-go result can be circumvented by breaking either
homogeneity or isotropy. In \cite{D'Amico:2011jj} inhomogeneous
cosmological solutions (see also 
\cite{Chamseddine:2011bu,Koyama:2011xz,Koyama:2011yg,Gratia:2012wt,Kobayashi:2012fz,Volkov:2012cf}
for related solutions) were considered and it was shown that the 
standard cosmological evolution in GR can be recovered by making the
Vainshtein radius as large as the size of the cosmological horizon. On
the other hand,  it was found in \cite{Gumrukcuoglu:2012aa} that the dRGT 
theory with the de Sitter or FLRW fiducial metric admits attractor solutions
with broken isotropy. On the attractor called anisotropic FLRW solution,
at the level of background, the anisotropy is entirely included in the
relation between physical and fiducial metrics and is thus  hidden from
observers probing the physical metric. The stability of anisotropic FLRW 
solutions was investigated in \cite{DeFelice:2013awa}.

An alternative possibility to evade the no-go result is to extend the
theory by introducing an extra degree(s) of freedom. For example, one can
promote each coupling constant in the dRGT action to a function of an
extra scalar field~\cite{D'Amico:2012zv,Huang:2012pe}. One can also
multiply the fiducial metric of the dRGT theory with a conformal factor that
is a function of an extra scalar field. The quasi-dilaton
theory~\cite{D'Amico:2012zv} is an example of this type and admits a
self-accelerating FLRW~\footnote{Here, FLRW indicates that both physical
and fiducial metrics are homogeneous and isotropic.} de Sitter
solution. While the self-accelerating de Sitter solution in the original
quasi-dilaton theory always exhibits ghost
instability~\cite{Gumrukcuoglu:2013nza,D'Amico:2013kya}~\footnote{In the
decoupling limit of the original quasi-dilaton, another type of
self-accelerating solution was recently found and claimed to be
stable~\cite{Gabadadze:2014kaa}.}, it is stable in
a regime of parameters in an extended version of the
quasi-dilaton, where not only conformal but also disformal
transformations are applied to the fiducial
metric~\cite{DeFelice:2013tsa,DeFelice:2013dua}.

A disformal transformation of the fiducial metric was first considered
in the context of massive gravity coupled to a DBI Galileon (DBI massive 
gravity)~\cite{Gabadadze:2012tr}. However, on all known homogeneous and
isotropic self-accelerating solutions in the DBI massive gravity, three
among the six degrees of freedom have vanishing (time) kinetic
terms~\cite{Andrews:2013uca,Goon:2014ywa} and  are thus expected to exhibit nonlinear instability as in the case of  dRGT. 

In this work, we explore yet another way to evade the no-go result and to have stable cosmological solutions in a massive gravity theory.
In cosmological applications of massive gravity theories the graviton mass is
often set to be of the order of the Hubble expansion rate $H$ of the
universe. Considering the fact that a non-vanishing $H$ spontaneously
breaks Lorentz invariance and respects $SO(3)$-invariance only, it is
natural to require that the graviton mass of order $H$ should be related
to $SO(3)$-invariance but not necessarily the full Lorentz
invariance. This consideration leads to yet another possibility to
extend the dRGT theory. Namely, we divide the $4$-dimensional fiducial
metric into a $3$-dimensional spatial fiducial metric and a $1$-form
corresponding to the derivative of the temporal St\"{u}ckelberg field,
and then treat them as independent ingredients in the action. This type
of massive gravity with $SO(3)$-invariance is not new and has been
considered in the
literature~\cite{Rubakov:2004eb,Dubovsky:2004sg,Blas:2009my}. 
A general $SO(3)$-invariant massive gravity with five degrees of freedom 
was recently constructed at fully nonlinear level in
\cite{Comelli:2013paa,Comelli:2013txa}. Cosmology in this 
general framework, with some restrictions, was then investigated in 
\cite{Comelli:2013tja} and it was shown that the (time)  kinetic terms
of the five degrees of freedom generically do not vanish on FLRW backgrounds
but that three among the five again have vanishing (time) kinetic terms on
de Sitter backgrounds.

The purpose of the present paper is to revisit cosmology in the $SO(3)$-invariant massive gravity of
\cite{Comelli:2013paa,Comelli:2013txa} in a setup that is
more general than in \cite{Comelli:2013tja}. Contrary to the case in
\cite{Comelli:2013tja}, we find that all five degrees of freedom can
have non-vanishing kinetic terms on strictly homogeneous and isotopic de
Sitter backgrounds. In particular, we provide a simple class of concrete
models that admit de Sitter solutions on which all $5$ degrees of
freedom are weakly coupled and stable.

The rest of the present paper is organized as follows.  
In the next section, we motivate and present the model, first in a covariant formulation, then in the unitary gauge. In section \ref{sec:3}, we consider  the homogeneous and isotropic solutions and derive the  Friedmann equations. In the subsequent section, section \ref{sec:perturbations}, we turn to the study of the linear perturbations around the homogeneous and isotropic solutions and derive the conditions for stability. In section \ref{sec:example}, we introduce a simple ansatz for the potential of massive gravity characterized by a few parameters  and apply to it the results of the previous sections. Section \ref{sec:comparison} is devoted to a brief comparison of our results with previous studies and we conclude in the final section.

\section{The model} 
\label{sec:2}

In the present paper we study cosmology in $SO(3)$-invariant massive
gravity. We first motivate and  describe the setup before going
into the details of the analysis in the following sections.

\subsection{Motivation}

The attempts to  modify gravity in the IR  can be divided, roughly speaking, into two categories.
One is to change the theory itself from  GR. A typical example in
this category is massive gravity, giving a non-vanishing mass to the
spin-$2$ graviton. The other is to change the state of the universe
without changing  GR itself. This is the idea of  Higgs phase of gravity,
and the simplest illustration is the ghost condensate~\cite{ArkaniHamed:2003uy}, in which
a Nambu-Goldstone boson associated with spontaneously broken time
diffeomorphism invariance is ``eaten'' by the graviton and modifies the IR
behavior of gravity. An advantage of the latter approach is that the
structure of the low-energy effective field theory is robustly
determined by the symmetry breaking pattern.

While massive gravity was recently promoted to a fully nonlinear
theory at least classically \cite{deRham:2010ik,deRham:2010kj},\footnote{However, see \cite{deRham:2013qqa}
for some evidences of technical
naturalness.} the regime of
validity of the effective field theory is very limited because of a
rather low cutoff scale $\Lambda_3\simeq (m_g^2M_{\rm Pl})^{1/3}$, 
where $m_g$ is the graviton mass. It is thus favorable to seek a
possible (partial) UV completion. Although a (partial) UV completion has
not yet been found, one ideal possibility may be to realize massive
gravity at low energy as a consequence of spontaneous symmetry breaking,
i.e. due to an analogue of the Higgs mechanism in gravity. This would
unify the two approaches to IR modification of gravity described above.

With this in mind, it seems natural to expect that the structure of the
low-energy effective field theory of massive gravity may depend on the
choice of background since different backgrounds generically have
different symmetry breaking patterns. A background of particular
interest is the cosmological one. Since a generic FLRW background
respects the $3$-dimensional maximal symmetry but breaks the
$4$-dimensional maximal symmetry, in the present paper we shall consider
a class of massive gravity theories which have the same symmetry
type. Furthermore we shall restrict our considerations to those with
five physical degrees of freedom, the same number of degrees of freedom
as a Lorentz-invariant massive spin-$2$ field in Minkowski
background. Fortunately, a general class of theories respecting the
$3$-dimensional maximal symmetry and with five degrees of freedom was
recently found in \cite{Comelli:2013paa,Comelli:2013txa}. In the present 
paper we thus investigate cosmology in this class of theories.

\subsection{Covariant description}

General massive gravity with  $3$-dimensional Euclidean symmetry has a
covariant description in terms of a $4$-dimensional physical metric
$g_{\mu\nu}$ and four scalar fields ($\Phi$, $\Phi^I$) ($I=1,2,3$). In
this covariant description, the action enjoys the $4$-dimensional
diffeomorphism invariance 
\begin{equation}
 x^{\mu} \to x'^{\mu}(x^{\nu}),
\end{equation}
as well as the internal $3$-dimensional Euclidean symmetry:
\begin{equation}
 \Phi^I \to \Phi^I + C^I, \quad
  \Phi^I \to V^I_{\, J}\Phi^J,
\end{equation}
where $C^I$ ($I=1,2,3$) are arbitrary constants and $V^I_{\, J}$ is an
arbitrary $SO(3)$ rotation. This global symmetry implies that the fields
$\Phi^I$ enter the action only through the fiducial metric $\f_{\mu\nu}$
defined by 
\begin{equation}
\f_{\mu\nu}
 \equiv \delta_{IJ}\, \partial_{\mu}\Phi^I\partial_{\nu}\Phi^J.
\end{equation}
Note that $\f_{\mu\nu}$ is a degenerate symmetric tensor from the
4-dimensional point of view.

On the other hand,  there is a priori no restriction on how $\Phi$
enters the action. Consequently, the action can be constructed from the
following ingredients:
\begin{equation}
 g_{\mu\nu}, \quad \nabla_{\mu}, \quad 
  R^{\mu}_{\ \nu\rho\sigma}, \quad
  \Phi,   \quad \f_{\mu\nu} \,,
\end{equation}
where $\nabla_{\mu}$ and $R^{\mu}_{\ \nu\rho\sigma}$ are, respectively, the covariant
derivative  and the Riemann curvature tensor associated with the metric $g_{\mu\nu}$. Since we are interested in a low-energy effective field
theory of gravity relevant to late-time cosmology, we shall not consider
higher derivative terms and restrict our attention to an action for
gravity of the following form (in the Einstein frame):
\begin{equation}
 S_{\rm grav} = M_{\rm Pl}^2 \int \dd^4x \sqrt{-g}
  \left[\frac{R}{2} - m^2
   V(g_{\mu\nu}, \Phi, \partial_{\mu}\Phi, \f_{\mu\nu})\right],
\label{action-general}
\end{equation} 
where $m$ ($>0$) corresponds to  the graviton mass, and
$V(g_{\mu\nu}, \Phi, \partial_{\mu}\Phi, \f_{\mu\nu})$ is a scalar
made of $g_{\mu\nu}$ (and its inverse $g^{\mu\nu}$), $\Phi$,
$\partial_{\mu}\Phi$ and $\f_{\mu\nu}$. 
Generically, a theory of this
form contains $6$ degrees of freedom, one of which being a BD ghost. In
order to obtain a ghost-free theory, we follow the work
\cite{Comelli:2013paa,Comelli:2013txa},  which constructed in a
systematic way potentials leading to ghost-free theories, with only five
physical degrees of freedom. We now summarize their results.

In order to write down the action of the theory with $5$
(instead of $6$) degrees of freedom, it is convenient to introduce a set
of $10$ scalars (${\cal N}$, ${\cal N}^I$, $\Gamma^{IJ}=\Gamma^{JI}$) as
\begin{equation}
 {\cal N} \equiv
  \frac{1}{\sqrt{-g^{\mu\nu}\partial_{\mu}\Phi\partial_{\nu}\Phi}}, 
  \quad
  {\cal N}^I \equiv {\cal N} n^{\mu}\partial_{\mu}\Phi^I, 
  \quad
  \Gamma^{IJ} \equiv (g^{\mu\nu}+n^{\mu}n^{\nu})
  \partial_{\mu}\Phi^I\partial_{\nu}\Phi^J,
\end{equation} 
where we have defined the unit vector 
\begin{equation}
 n^{\mu} \equiv {\cal N} g^{\mu\nu}\partial_{\nu}\Phi.
\end{equation}
We also introduce a set of $3$ auxiliary scalars $\xi^I$ ($I=1,2,3$) and then
define $6$ additional  scalars ${\cal K}^{IJ}$ ($={\cal K}^{JI}$)
($I,J=1,2,3$) by 
\begin{equation}
 {\cal K}^{IJ} \equiv \Gamma^{IJ}-\xi^I\xi^J. 
\end{equation} 
The theory is then characterized by two arbitrary $SO(3)$-invariant
functions
\begin{equation}
 {\cal U}({\cal K}^{IJ},\delta_{IJ},\Phi)\quad \mbox{and}\quad
  {\cal E}(\Gamma^{IJ},\xi^I,\delta_{IJ},\Phi). 
\end{equation}
To write down the potential $V$
explicitly, we first impose the following three constraints relating
$\xi^I$ to ${\cal N}^I$ ($I=1,2,3$):
\begin{equation}
 {\cal N}^I - {\cal N}\xi^I = - {\cal U}^{IJ}{\cal E}_J, \label{eqn:def-xi}
\end{equation}
where ${\cal U}^{IJ}$ is the inverse of the Hessian matrix 
${\cal U}_{IJ}\equiv\partial^2{\cal U}/\partial\xi^I\partial\xi^J$ and
${\cal E}_J\equiv\partial{\cal E}/\partial\xi^J$. The potential $V$ is
then specified as
\begin{equation}
 V = {\cal U} + \frac{{\cal E}-{\cal U}_I\, {\cal U}^{IJ}\, {\cal E}_J}{\cal N},
  \label{eqn:potential-general}
\end{equation}
where ${\cal U}_I\equiv\partial{\cal U}/\partial\xi^I$.
It has been shown in the unitary gauge that the form of the potential term $V$, specified by (\ref{eqn:potential-general}), guarantees to suppress one of the original $6$ degrees of freedom, leaving only $5$ physical degrees of freedom in the theory \cite{Comelli:2013paa,Comelli:2013txa}. This will also be manifest in our linear analysis (and discussed in particular in Appendix~\ref{app:noBD}).

\subsection{Fiducial metric}
\label{subsec:fiducial}

The general theory is characterized by two $SO(3)$-invariant functions 
${\cal U}({\cal K}^{IJ},\delta_{IJ},\Phi)$ and 
${\cal E}(\Gamma^{IJ},\xi^I,\delta_{IJ},\Phi)$. While the
$SO(3)$-invariance restricts the way these two functions depend on
$\Gamma^{IJ}$ and $\xi^I$, the dependence on $\Phi$ is arbitrary. This
arbitrariness significantly reduces the predictability of the theory. In
the following we thus consider two specific cases with additional global 
symmetries that restrict the way $\Phi$ can enter 
${\cal U}({\cal K}^{IJ},\delta_{IJ},\Phi)$ and 
${\cal E}(\Gamma^{IJ},\xi^I,\delta_{IJ},\Phi)$. Motivated by 
this, we then propose a prescription that is general enough
to cover the two cases specified by symmetries and that is still simple
enough for the analysis in the forthcoming subsections.

\subsubsection{Case with shift symmetry}

One of the simplest symmetries one can envisage is the shift symmetry, i.e. 
\begin{equation}
 \Phi \to \Phi + C,
\end{equation} 
where $C$ is an arbitrary constant. In this case, the two functions are of the form
\begin{equation}
{\cal U}={\cal U}({\cal K}^{IJ},\delta_{IJ}), \quad
{\cal E}={\cal E}(\Gamma^{IJ},\xi^I,\delta_{IJ}).
\end{equation} 
Theories of this type however do not admit FLRW solutions, unless the
function ${\cal E}$ is fine-tuned \cite{Comelli:2013tja}.
 (As a special case, this class of theories includes
 the original dRGT theory with Minkowski fiducial metric,
 which does not allow for the FLRW cosmology.) 
Therefore, we will not explore further this class of theories.

\subsubsection{Case with dilaton-like symmetry}

As a simple deformation of the shift symmetry,  we now consider the
following global symmetry:
\begin{equation}
\label{sym}
 \Phi \to \Phi + C, \quad \Phi^I \to e^{-MC}\Phi^I,
\end{equation} 
where $C$ is an arbitrary constant and $M$  some energy scale. In this
case, the two functions ${\cal U}$ and ${\cal E}$ are of the form 
\begin{equation}
{\cal U}=\tilde{\cal U}(\tilde{\cal K}^{IJ},\delta_{IJ}), \quad
{\cal E}=\tilde{\cal E}(\tilde{\Gamma}^{IJ},\tilde{\xi}^I,\delta_{IJ}),
\end{equation} 
where we have introduced the rescaled quantities
\begin{equation}
 \tilde{\Gamma}^{IJ} \equiv e^{2M\Phi}\Gamma^{IJ}, \quad
  \tilde{\xi}^I \equiv e^{M\Phi}\xi^I, \quad
  \tilde{\cal K}^{IJ} \equiv e^{2M\Phi}{\cal K}^{IJ}
  = \tilde{\Gamma}^{IJ}-\tilde{\xi}^I\tilde{\xi}^J\,,
\end{equation}
which are invariant under the transformation (\ref{sym}).

The constraints (\ref{eqn:def-xi}) and the potential
(\ref{eqn:potential-general}) are expressed in terms of them as
\begin{equation}
 \tilde{\cal N}^I - {\cal N}\tilde{\xi}^I = - \tilde{\cal U}^{IJ}\, \tilde{\cal E}_J, 
\end{equation}
and
\begin{equation}
 V = \tilde{\cal U} + \frac{\tilde{\cal E}-\tilde{\cal U}_I\, \tilde{\cal
  U}^{IJ}\, \tilde{\cal E}_J}{\cal N},
\end{equation}
where $\tilde{\cal U}^{IJ}$ is the inverse of the Hessian matrix 
$\tilde{\cal U}_{IJ}\equiv\partial^2\tilde{\cal U}/\partial\tilde{\xi}^I\partial\tilde{\xi}^J$, 
$\tilde{\cal U}_I\equiv\partial\tilde{\cal U}/\partial\tilde{\xi}^I$, 
$\tilde{\cal E}_J\equiv\partial\tilde{\cal E}/\partial\tilde{\xi}^J$, and 
$\tilde{\cal N}^I\equiv e^{M\Phi}{\cal N}^I$.

\subsubsection{Prescription with general fiducial metric}

Motivated by the two specific cases considered above, we now propose a
class of theories specified by the two $SO(3)$-invariant functions
${\cal U}$ and ${\cal E}$ of the form 
\begin{equation}
{\cal U}={\cal U}({\cal K}^{IJ},f_{IJ}), \quad
{\cal E}={\cal E}({\Gamma}^{IJ},\xi^I,f_{IJ}), \label{eqn:UE-ansatz}
\end{equation} 
where 
\beq
f_{IJ}=b^2(\Phi)\delta_{IJ}\,, \label{eqn:3dfiducialmetric}
\eeq
is the fiducial three-dimensional metric including the dependence on the
scalar field $\Phi$. The auxiliary scalars $\xi^I$ and the potential $V$
are then constructed as in (\ref{eqn:def-xi}) and
(\ref{eqn:potential-general}).

This prescription includes the two particular cases discussed above: 
$b(\Phi)=1$ for the case with shift symmetry; and 
$b(\Phi)=e^{M\Phi}$ for the case with dilaton-like symmetry. In the
forthcoming sections, we treat $b(\Phi)$ as a general positive function
of $\Phi$, which can be seen as a scale factor in the 3-d field space
spanned by the $\Phi^I$, parametrized by $\Phi$.

\subsection{ADM formulation in the unitary gauge}

For simplicity, it is  convenient to work directly in the
unitary gauge, i.e. in the privileged coordinate system associated with
the preferred slicing of the theory, such that 
\beq
\Phi = t, \quad \Phi^i = x^i\,,
\eeq
and to write the metric in the ADM
form 
\beq
 \dd s^2=g_{\mu\nu} \dd x^\mu \dd x^\nu
 = -N^2 \dd t^2 +\gamma_{ij} (\dd x^i+N^i \dd t) ( \dd x^j+N^j \dd t ) \,,
\label{ADM-line}
\eeq
where $N$ denotes the lapse function and $N^i$ the shift vector. 
In this framework, the theory with general fiducial metric is specified
by the two scalars 
\begin{equation}
{\cal U}={\cal U}({\cal K}^{ij},f_{ij}), \quad
{\cal E}={\cal E}(\gamma^{ij},\xi^i,f_{ij}), 
\end{equation} 
where $\gamma^{ij}$ is the inverse of the spatial metric $\gamma_{ij}$, 
\begin{equation}
 f_{ij} = b^2(t)\delta_{ij}\,,
\end{equation}
is the fiducial three-dimensional metric in the unitary gauge, and the
three-dimensional vector $\xi^i$ and the three-dimensional symmetric
tensor ${\cal K}^{ij}$ are defined through
\begin{equation}
\label{xi}
 N^i - N \xi^i = - {\cal U}^{ij}{\cal E}_j, 
  \quad 
{\cal K}^{ij} = \gamma^{ij} - \xi^i\xi^j.
\end{equation}
Here, ${\cal U}^{ij}$ is the inverse of the Hessian matrix 
${\cal U}_{ij}\equiv\partial^2{\cal U}/\partial\xi^i\partial\xi^j$, and
${\cal E}_j\equiv\partial{\cal E}/\partial\xi^j$. The potential $V$ in
the unitary gauge is then specified as
\begin{equation}
 V\left(N, N^i,\gamma^{ij}, f_{ij}\right) = 
  {\cal U} + \frac{{\cal E}-{\cal U}_i\, {\cal U}^{ij}\, {\cal E}_j}{N},
\end{equation}
where ${\cal U}_i\equiv\partial{\cal U}/\partial\xi^i$. The action in
the gravity sector is then of the form 
\beq
S_{\rm grav}=S_{\rm EH}+S_{\rm mg} =
 \frac{M_{\rm Pl}^2}{2}
 \int \dd^4 x N\sqrt{\gamma}\left(K_{ij} K^{ij}-K^2+{}^{(3)}R\right)
 -M_{\rm Pl}^2m^2 
 \int \dd^4x N \sqrt{\gamma}\, V\left(N, N^i,\gamma^{ij},f_{ij}\right) \,,
\eeq
where  $\gamma\equiv\det(\gamma_{ij})$, and ``EH'' and ``mg'' denote the Einstein-Hilbert and massive gravity terms, respectively. 
The first part of the action corresponds to the ADM expression of 
the Einstein-Hilbert action, where
\beq
K_{ij}\equiv\frac{1}{2N}\left( \dot{\gamma}_{ij}-N_{i|j}-N_{j|i}\right)\,, \qquad K\equiv \gamma^{ij}K_{ij}
\eeq
are the extrinsic curvature tensor and its trace
(the symbol $|$ denotes the spatial covariant derivative associated with 
the spatial metric $\gamma_{ij}$), while ${}^{(3)}R$ denotes the
three-dimensional Ricci scalar associated with the metric
$\gamma_{ij}$. The second part of the action consists of  the massive
gravity potential in the unitary gauge, which in general depends
explicitly on the time $t$ through the fiducial metric $f_{ij}$.

\section{Homogeneous cosmology}
\label{sec:3}

Let us now consider the cosmology of this class of theories. It is natural to
assume that the preferred frame associated with the massive gravity
potential  coincides with the cosmological frame (i.e. comoving
observers are at rest in the preferred frame). Assuming a spatially flat
FLRW spacetime, the physical metric is thus given by 
\beq
\label{metric_FLRW}
 \dd s^2=g_{\mu\nu} \dd x^\mu \dd x^\nu=
-\bar{N}^2(t) \dd t^2 +a^2(t) \delta_{ij} \dd x^i \dd x^j\,,
\eeq
i.e. 
\beq
\bar{N} = \bar{N}(t), \quad \bar{N}^i=\bar{\xi}^i=0, \quad 
\bar{\gamma}_{ij} = a^2(t)\delta_{ij}\,. 
\eeq

In the background, the potential reduces to 
\beq
\bar{V}=\bar{{\cal U}}+\frac{\bar{{\cal E}}}{\bar{N}}\,,
\eeq
where an overbar denotes that the corresponding quantity is evaluated on
the background. Adding the matter action $S_m$, the total action on the
background is thus given by 
\beq
\bar S_{\rm total}=\bar S_{\rm grav}+\bar S_m=
M_{\rm Pl}^2 \int \dd^4x\, a^3
\left( -3\frac{\dot a^2}{\bar{N}a^2}
-m^2 \bar{N}\, \bar{{\cal U}} -m^2 \bar{{\cal E}}\right)
+\bar S_m\,,
\eeq
where an overdot represents a derivative with respect to the time $t$. 

Both $\bar{{\cal U}}$ and $\bar{{\cal E}}$ depend on the scale factors $a$ and $b$, 
via their dependence on $\gamma^{ij}$ and  $f_{ij}$ respectively. 
More precisely, they  depend on the
ratio 
\beq
\label{X}
X\equiv \frac{b}{a}\,,
\eeq
since $\bar{{\cal U}}$, like $\bar{{\cal E}}$, depends on scalar combinations of the
matrix 
\beq
\bar \gamma^{ik}f_{kj}= \frac{b^2}{a^2}\, \delta^i_j=X^2\, \delta^i_j\,.
\eeq
where $\bar{\gamma}^{ij}=\delta^{ij}/a^2(t)$ is the inverse of
$\bar{\gamma}_{ij}=a^2(t)\delta_{ij}$.
Introducing the notations 
\beq
\label{V'}
\overline{\left(\frac{\p\U}{\p\K^{ij}}\right)}\equiv \U' \bar\g_{ij}\,, \qquad
\overline{\left(\frac{\p\E}{\p\g^{ij}}\right)}\equiv \E' \bar\g_{ij}\,, 
\eeq
one immediately finds
\beq
\label{U_X}
X \frac{\p\bar\U}{\p X}=6\U', \qquad X \frac{\p\bar\E}{\p X}=6\E'\,.
\eeq
This implies in particular $a\partial \bar{\cal U} / \partial a = - 6 {\cal U}' $ and $a\partial \bar{\cal E} / \partial a = - 6 {\cal U}' $.

The variation of the total homogeneous action $\bar{S}_{\rm total}$, with respect to
$\bar{N}$ and to $a$, yields the  Friedmann equations, which read
\beq
 3M_{\rm Pl}^2H^2=\rho_g+\rho_m, \qquad  
M_{\rm Pl}^2\left(2\frac{\dot{H}}{\bar{N}}+3H^2\right)=-P_g-P_m\,,
 \label{Friedman-eq}
\eeq
where $H\equiv \dot{a}/(\bar{N}a)$ is the Hubble expansion rate. On the right hand sides, we find the matter energy density and pressure, respectively denoted by 
$\rho_m$ and $P_m$,  as well as the effective energy density and pressure associated with the massive gravity potential, expressed as 
\beq
 \rho_g\equiv M_{\rm Pl}^2m^2 \bar{{\cal U}}, \qquad  
 P_g\equiv M_{\rm Pl}^2m^2
 \left[2\U'-\bar{{\cal U}}
 +\frac{1}{\bar{N}}\left(2\E'-\bar{{\cal E}}\right)\right]\,.
 \label{eqn:rhog-Pg}
\eeq
Even in the absence of matter, one can obtain an accelerating solution
$\partial_t(\dot{a}/\bar{N})>0$, provided 
\beq
\rho_g+3P_g=M_{\rm Pl}^2m^2
\left[6\U'-2\bar{{\cal U}}
+\frac{1}{\bar{N}}\left(6\E'-3\bar{{\cal E}}\right)\right] <0\,.
\eeq

The Bianchi identity and the conservation of ordinary matter imply the following conservation equation for the effective gravitational component: 
 \beq
 \label{conserv_mg}
\dot \rho_g+3\bar{N}H\left(\rho_g+P_g\right)=
M_{\rm Pl}^2m^2\left[ \dot{\bar{{\cal U}}}+3H\left(2\bar{N}\U'+2\E'-\bar{{\cal E}}\right)\right]=0\,.
\eeq
The quantity $\bar{{\cal U}}$ is in general time-dependent, via its dependence on $X$, and 
using (\ref{U_X}), one finds
\beq
\dot{\bar{{\cal U}}} =-6\left(\bar{N}H-\frac{\dot{b}}{b}\right)\U'\,.
\label{Udot2Up}
\eeq
Substituting this into (\ref{conserv_mg}), one finally obtains the 
relation,\footnote{This relation can also be obtained directly from the
action by keeping the time-like St\"{u}ckelberg field $\Phi$ 
and taking the  variation of the action with respect to
$\Phi$.
In this case, $\dot{b}$ is replaced by $\partial_\Phi b$.
} 
\beq
\U' \frac{\dot b}{b}+H\left(\E'-\frac12\bar{{\cal E}}\right)=0\,.
\label{eqn:constraint}
\eeq
This constraint equation complements the Friedmann equations given in (\ref{Friedman-eq}).

In the absence of matter ($\rho_m=P_m=0$), and assuming expansion
($H>0$), we have from (\ref{Friedman-eq}) and (\ref{eqn:constraint}), 
\begin{equation}
 H = \frac{m}{\sqrt{3}}\sqrt{\bar{{\cal U}}}\,,\quad 
  {\cal U}' \frac{\dot b}{b} + \frac{m}{\sqrt{3}}\sqrt{\bar{{\cal U}}}
   \left({\cal E}'-\frac12\bar{{\cal E}}\right)=0\,.
  \label{eqn:vaccum-background}
\end{equation}
Since the ratio $\dot{b}/b$ is not dynamical but fixed by the theory,
the second equation above should be considered as an algebraic equation
for $X$. Once this is solved with respect to $X$, the first equation
determines the background geometry.

For example, in the theory with dilaton-like symmetry considered in
subsection \ref{subsec:fiducial}, we have $\dot{b}/b=M$. Hence we obtain
\begin{equation}
  {\cal U}' M + \frac{m}{\sqrt{3}}\sqrt{\bar{{\cal U}}}
   \left({\cal E}'-\frac12\bar{{\cal E}}\right)=0\,,
   \label{eqn:eq-X-vacuum}
\end{equation}
which is a time-independent algebraic equation for $X$. Provided
that this equation allows for a real solution $X=X_0$, the Hubble
expansion rate is determined by 
\begin{equation}
 H = \frac{m}{\sqrt{3}}\sqrt{\bar{{\cal U}}|_{X=X_0}}\,.
  \label{eqn:H-vacuum}
\end{equation} 
Hence the graviton mass term acts as an effective cosmological constant,
implying that all vacuum solutions are de Sitter 
once we impose the dilaton-like symmetry (\ref{sym}).
In this example the value of the lapse $\bar{N}$ is set by the model parameters through $\bar{N} = M / H$, assuming ${\cal U}' \ne 0$, so that the coordinate time $t$ does not correspond to the usual intuitive cosmic time, i.e. the proper time $\tau=\int \bar{N}\dd t$. As seen from (\ref{eqn:eq-X-vacuum}) and (\ref{eqn:H-vacuum}), changing $M$ implies changing the value of $X_0$ and thus that of $H$ in a de Sitter vacuum. In this sense, the parameter $M$, i.e. the dependence of $b$ on $\Phi$, directly influences the relation between $H$ and $X$.

\section{Linear perturbations}
\label{sec:perturbations}

We now turn to the study of linear perturbations about the homogeneous 
solutions introduced in the previous section. For simplicity we consider
the pure massive gravity system without matter.

Perturbations of the background metric (\ref{metric_FLRW}), in the ADM
form (\ref{ADM-line}), are described by the lapse perturbation $\delta
N$, defined by 
\begin{equation}
 N = \bar{N}(t) + \delta N\,,
\end{equation} 
the shift vector $N^i$, which is already a perturbation as it vanishes
in the background, and the perturbations $h_{ij}$ of the
three-dimensional metric, defined by 
\beq
\gamma_{ij}=\bar{\gamma}_{ij}+\delta\gamma_{ij}
=a^2(t)\left(\delta_{ij}+h_{ij}\right)\,.
\eeq
We expand the total gravitational action up to quadratic order in the
perturbations. As shown in Appendix~\ref{app:EHaction}, the quadratic 
part of the Einstein-Hilbert action can be written as 
\begin{equation}
S_{\rm EH}^{(2)} = \tilde{S}_{\rm EH}^{(2)}
 + M_{\rm Pl}^2\int \dd^4x \left( N\sqrt{\gamma} \,\right)^{(2)} 3H^2\,,
\end{equation} 
where $\left( N\sqrt{\gamma} \,\right)^{(2)}$ is the quadratic part of
$N\sqrt{\gamma}$ and the explicit expression for $\tilde{S}_{EH}^{(2)}$
is given in (\ref{action_EH_quad}).

On the other hand, the graviton mass term, expanded up to quadratic
order, yields 
\bea
S_{\rm mg}&=&-M_{\rm Pl}^2m^2 \int \dd^4x 
\Bigl( N\sqrt{\gamma}\,\bar{{\cal U}}+\sqrt{\g}\bar\E \Bigr)
\cr
&& -M_{\rm Pl}^2m^2 \int \dd^4x \, \sqrt{\g}\left\{\left(N \U' +\E'\right)\bar\g_{ij}\d \g^{ij}
-\left(\bar{N}\U'+\frac12 \E''\right)\bar\g_{ij}\xi^i\xi^j
\right.
\cr 
&& \left.
+\left[\frac12 \left(\bar{N}\U_s''+\E_s''\right)\bar\g_{ij}\bar\g_{kl}+\frac14\left(\bar{N}\U_t''+\E_t''\right)
\left(\bar\g_{ik}\bar\g_{jl}+\bar\g_{il}\bar\g_{jk}\right)\right]\d\g^{ij}\d \g^{kl}
\right\} + O(\epsilon^3)\,,
\eea
where $\epsilon$ counts the order of the perturbative expansion, and the
``second'' derivatives of $\E$ and $\U$ are defined as follows: 
\bea
 \overline{\left(\frac{\p^2\U}{\p\K^{ij}\p \K^{kl}}\right)}
 &\equiv& \U''_s \bar\g_{ij} \bar\g_{kl}
+\frac12 \U''_t \left(\bar\g_{ik} \bar\g_{jl} + \bar\g_{il} \bar\g_{jk}\right)\,,
\label{V''1}
\\
\overline{\left(\frac{\p^2\E}{\p\g^{ij}\p \g^{kl}}\right)}
&\equiv& \E''_s \bar\g_{ij} \bar\g_{kl}
+\frac12 \E''_t \left(\bar\g_{ik} \bar\g_{jl} + \bar\g_{il} \bar\g_{jk}\right)\,, \qquad 
\overline{\left(\frac{\p^2\E}{\p\xi^i\p\xi^j}\right)}\equiv \E'' \bar\g_{ij}\,.
\label{V''2}
\eea
Here, we recall that  quantities with overbar are those evaluated on the background. Using 
\begin{equation}
\xi^i=\frac{2\U'}{2\bar{N}\U'+\E''}N^i+{\cal O}(\epsilon^2)\,, 
\eeq
which follows from (\ref{xi}), and 
\beq
\gamma^{ij}=
\frac{1}{a^2}\left(\d^{ij}-h^{ij}+h^{ik}h_{k}^{\; j}\right)
+{\cal O}(\epsilon^3)\,,
\end{equation}
the graviton mass term, up to quadratic order, is rewritten in the form 
\bea
\label{S_mg_quad}
S_{\rm mg}&=& - \int \dd^4x N\sqrt{\gamma}\,\rho_g 
-M_{\rm Pl}^2m^2 \int \dd^4x \, a^3\left\{\bar\E+\left(\frac12\E-\bar{N}\U'-\E'\right)h
\right.
\cr 
&& \left.
-\U' \, \d N h-\frac{2\U'^2}{2\bar{N}\U'+\E''} a^2\delta_{ij}N^i N^j+\frac12\left[\frac{\bar\E}{4}-\bar{N}\U'-\E'+\bar{N}\U_s''+\E_s''\right]h^2
\right.
\cr 
&& \left.
+\left[-\frac{\E}{4}+\bar{N}\U'+\E'+\frac12\left(\bar{N}\U_t''+\E_t''\right)\right]h_{ij}h^{ij}
\right\}\ + O(\epsilon^3)\,,
\eea
where $\rho_g$ is defined in (\ref{eqn:rhog-Pg}), $h\equiv \delta^{ij}h_{ij}$ and
hereafter all spatial indices are raised and lowered by $\delta^{ij}$
and $\delta_{ij}$,
respectively. This leads to the following expression for the
quadratic part of the graviton mass term:
\beq
S_{\rm mg}^{(2)}  =  \tilde{S}_{\rm mg}^{(2)} 
- \int \dd^4x \left( N\sqrt{\gamma} \, \right)^{(2)} \rho_g\,, 
\eeq
with
\beq
\tilde{S}_{\rm mg}^{(2)}  \equiv  
 M_{\rm Pl}^2 m^2 \int \dd^4x \, \bar{N}a^3
 \left( c_0\, \frac{\d N^2}{\bar{N}^2}+c_1\, \frac{\d N}{\bar{N}} h
  + c_2 \bar \g_{ij}\frac{N^i N^j}{\bar{N}^2}
  +c_3\, h^2+c_4 \, h_{ij} h^{ij}\right)\,,
 \label{eqn:Smg-general}
\eeq
where the coefficients in the expansion are given by
\bea
c_0&=&0\,,\qquad c_1=\U'\,,\quad c_2=\frac{2\bar{N}\U'^2}{2\bar{N}\U'+\E''}\,, \quad c_3=-\frac12\left(\frac{\bar\E}{4\bar{N}}-\U'-\frac{\E'}{\bar{N}}+\U_s''+\frac{\E_s''}{\bar{N}}\right)
\cr
 c_4&=&\frac{\bar\E}{4\bar{N}}-\U'-\frac{\E'}{\bar{N}}-\frac12\left(\U_t''+\frac{\E_t''}{\bar{N}}\right)\,.
\label{ci-explicit}
\eea
We have explicitly added a term proportional to $\delta N^2$ in
(\ref{eqn:Smg-general}), even if this term vanishes in the present case,
in order to show that a generic potential  will generate  all the terms 
listed in the quadratic action above. As discussed in Appendix
\ref{app:noBD}, the vanishing of $c_0$ is directly related to the
absence of the sixth degree of freedom, i.e. BD ghost, in the models
under study. 

We can then rewrite the quadratic action for the pure massive gravity
system as 
\begin{equation}
S_{\rm grav}^{(2)} = \tilde{S}^{(2)}_{\rm EH} + \tilde{S}^{(2)}_{\rm mg}
 + \int \dd^4x
 \left( N\sqrt{\gamma} \, \right)^{(2)} \left( 3 M_{\rm Pl}^2 H^2 - \rho_g \right) \; .
\label{action-tot}
\end{equation}
The last term in (\ref{action-tot}) vanishes once the background
equation is imposed, and  it thus makes no contribution to the analysis
of perturbations. As usual, it is convenient to decompose the
perturbations into scalar, vector and tensor modes. We discuss in turn
each of these sectors in the following subsections.

\subsection{Tensor modes}
\label{subsec:tensor}

Tensor modes are characterized by $\d N=0$, $N^i=0$ and the transverse
traceless condition on $h_{ij}$:
\beq
\p^ih_{ij} =0,\qquad h=0\,.
\label{tensor-dec}
\eeq
We then find that the quadratic action for the tensor modes is given by 
\bea
 S_T^{(2)}&=& M_{\rm Pl}^2\int \dd^4x \, \bar{N}a^3 \left( \frac18 \frac{\dot h_{i j}  \dot h^{ij}}{\bar{N}^2}-\frac{1}{8a^2} \, \p_k h_{ij} \, \p^k h^{ij}-\frac18 m_T^2 h_{ij}h^{ij}
\right) \,,
\label{action-tensor}
\eea
with 
\beq
\label{m_T}
m_T^2
= 4\frac{\dot{H}}{\bar{N}}-8 m^2 c_4
= -\frac{2}{M_{\rm Pl}^2}(\rho_g+P_g)-8 m^2 c_4\,,
\eeq
where we have used the background equations of motion, (\ref{Friedman-eq})-(\ref{eqn:rhog-Pg}), in absence of matter (i.e. with $\rho_m=P_m=0$), to obtain the last equality. As is clear from (\ref{action-tensor}), the tensor modes always have a healthy kinetic term, and the high-energy limit of their propagation speed is unity.

\subsection{Vector modes}
\label{subsec:vector}

For vector perturbations, the lapse has no vector mode, i.e. $\delta N = 0$, the shift is restricted to be transverse,
i.e. $\p_i N^i=0$, and the metric is decomposed as
\beq
h_{ij}=\p_i E_j+\p_j E_i,\qquad \p_i E^i=0\,.
\label{vector-dec}
\eeq
The quadratic action for vector modes reads
\begin{equation}
 S_V^{(2)} = M_{\rm Pl}^2\int \dd^4x \, \frac{a^3}{\bar{N}}\left( \frac14 \p_i \dot E_j \p^i \dot E^j+\frac14\p_iN_j\p^iN^j-\frac12 \p_i\dot E_j \p^iN^j
 -\frac14 \bar{N}^2m_T^2\p_iE_j \p^iE^j+\mu^2a^2 \delta_{ij} N^i N^j
 \right),
\end{equation}
with 
\beq 
 \mu^2\equiv c_2m^2 \,.
\label{eqn:mu-def}
\eeq
The shift components $N^i$ are not dynamical and can be integrated
out as follows. The momentum constraint, obtained by varying the action
with respect to the shift, yields the relation
\beq
 N_i=\frac{k^2}{k^2+4\mu^2a^2}\dot E_i\,,
 \label{shift-vec-DL}
\eeq
in Fourier space. Substituting this back into the action, the final action for
the two dynamical modes $E_i$ is given by 
\beq
 S_V^{(2)}=M_{\rm Pl}^2\int \dd t\, d^3k \,\bar{N} a^3 \left( \frac{k^2\mu^2a^2}{k^2+4\mu^2a^2}\frac{\vert \dot E_i \vert^2}{\bar{N}^2} -\frac{k^2}{4} m_T^2 \vert E_i \vert^2\right)
 \,.
 \label{action-vector}
\eeq
Note that the coefficient of the kinetic term vanishes for $\mu=0$,
i.e. for ${\cal U}'=0$ according to (\ref{ci-explicit}) and
(\ref{eqn:mu-def}),  which suggests  a strong coupling problem in this
limit.\footnote{The kinetic term vanishes when ${\cal U}' = 0$, not only
for the vector sector but also for the scalar sector, as will be seen in
the following subsection. In fact, when one considers the limit in which
the present theory becomes  equivalent to the dRGT theory, the so-called
self-accelerating branch corresponds to the limit ${\cal U}' = 0$. 
It has been shown that this branch of the dRGT theory suffers from a
nonlinear ghost due to the vanishing (time) kinetic terms of three out
of the five degrees of freedom \cite{DeFelice:2012mx} and our result for
${\cal U}' = 0$ is thus consistent with this. Moreover, seen from
(\ref{Udot2Up}), in the case of a trivial fiducial metric $\dot{b} = 0$,
a pure de Sitter solution ${\dot{\bar{\cal U}}} = 0$ necessarily
requires ${\cal U}' = 0$, leading to the vanishing of the kinetic
terms. This is also consistent  with the results of
\cite{Comelli:2013tja}. A more elaborate discussion on this issue is
given  in Section \ref{sec:comparison}.}
In the following, in order to avoid this problem as well as  ghost-like
pathologies related to a negative kinetic sign, we will require $\mu^2$
to be strictly positive, which is equivalent to the condition 
\beq
 c_2>0\,. \label{eqn:noghost-vector}
\eeq

The propagation speed for the vector modes can be read from the action
(\ref{action-vector}), by taking the ratio between the coefficient of
the gradient term and that of the kinetic term, which yields 
\beq
 c_V^2=\frac{k^2+4\mu^2a^2}{4k^2\mu^2}m_T^2\,.
 \eeq
In the high momentum limit, $k^2/a^2\gg \mu^2$, this reduces to 
 \beq
 c_V^2\simeq \frac{m_T^2}{4\mu^2}\qquad (k^2/a^2\gg \mu^2) \,.
 \eeq
One thus concludes that, under the no-ghost condition
(\ref{eqn:noghost-vector}), there is no gradient instability in the
vector sector if and only if $m_T^2>0$, 
when matter is negligible.

Before ending this subsection, let us comment on the Minkowski limit ${\cal H} \equiv H / M \to 0$ in the theory with dilaton-like symmetry (\ref{sym}), $b (\Phi) = e^{M \Phi}$. By imposing this symmetry, it follows from
 (\ref{Udot2Up}) that $\bar{N} = 1 / {\cal H}$, since $\dot{\bar{{\cal U}}}=0$ and $\dot b/b=M$, and from (\ref{eqn:constraint}) that ${\cal U}' \sim {\cal O}({\cal H})$, if $\bar{\cal E} \sim {\cal E}' \sim {\cal O}(1)$. Further assuming that the other derivatives of ${\cal U}$ are of order ${\cal O}({\cal H})$ and that those of ${\cal E}$ are of order ${\cal O}(1)$, we find that all $c_i$ are of ordrer ${\cal O}({\cal H})$, and thus $\mu^2 \sim m_T^2 \sim m^2 \times {\cal O}({\cal H})$. Therefore in the limit ${\cal H} \to 0$ the vector modes do not suffer from a pathological, i.e. either infinite or vanishing, propagation speed, which would typically be an indicator of strong coupling at nonlinear level.

\subsection{Scalar modes}
\label{subsec:scalar}

Scalar perturbations are described by $\d N$, as well as 
\beq
N^i=\delta^{ij}\p_j B, \qquad h_{ij}=2C \d_{ij}+2\p_i\p_j E\,.
\label{scalar-dec}
\eeq

The quadratic gravitational action thus depends on the four quantities
$\d N$, $B$, $C$ and $E$. The quantities $\delta N$ and $B$ are
manifestly non-dynamical and variation of the action with respect to
them yields two constraints, corresponding respectively to the
Hamiltonian constraint and the momentum constraint. From these two
constraints, one can extract the expressions of $\delta N$ and $B$ in
terms of the variables $E$ and $C$, and substitute them back into the
action. This gives an action that depends only on $C$ and $E$.

In the case $c_0=0$, it
turns out that the variable $C$ 
  is also non-dynamical and can  be integrated out,
 One finally obtains an action in terms of the only degree of freedom left,
 of the form
\beq
 S_S^{(2)}= M_{\rm Pl}^2\int \dd t\, \dd^3k \,\bar{N} a^3\left( {\cal K} \frac{\vert \dot E \vert^2}{\bar{N}^2}+{\cal M} \vert E \vert^2\right) \,.
 \label{action-scalar}
\eeq 
For simplicity, let us now consider a de Sitter background solution,
with constant $H$. According to (\ref{eqn:vaccum-background}), $X$ must
then be constant, as well as  $\dot{b}/b$, which corresponds to a model
with  dilaton-like symmetry. Since $\bar{N}=(\dot{b}/b)/H$ is also
constant, we conclude that all the coefficients defined  in
(\ref{ci-explicit}) are time-independent. In this case, the coefficient
of the kinetic term reduces to 
\beq
 \calK
 = \frac{m^2 k^4}{{\cal D}_E}\left\{ c_1 ( - c_1 + 2 c_2 ) \tilde k^2
 + 3 c_2 \Bigl[ c_1^2 m^2
 + \left( - 3 c_1 + 12 c_3 + 4 c_4 \right) H^2 \Bigr] \right\}\,,
 \label{eqn:calK-def}
\eeq
where $\tilde k\equiv k/a$ and
\beq
{\cal D}_E \equiv c_2 \tilde k^4 + 3 \Bigl[2 c_1 c_2 m^2  + \left( - 3 c_1 + 12 c_3 + 4 c_4 \right) H^2 \Bigr] \tilde k^2
+ 9  c_2 m^2 \Bigl[ c_1^2 m^2 + \left( - 3 c_1 + 12 c_3 + 4 c_4 \right) H^2
 \Bigr] \,.
\label{eqn:DE-def}
 \eeq
The coefficient of the mass term in the same de Sitter background is
given by 
\beq
{\cal M}=\frac{m^2 k^4}{{\cal D}_E^2} \Bigl( {\cal M}_8 \tk^8+{\cal M}_6 \tk^6+{\cal M}_4 \tk^4+{\cal M}_2 \tk^2 + {\cal M}_0 \Bigr),
\eeq
with
\bea
{\cal M}_8&=&4 c_2^2 (c_3 + c_4)\,,
\\
{\cal M}_6&=& c_2 \left( 3 c_1 - 12 c_3 - 4 c_4 \right)  \left(2 c_1 - 5 c_2 - 12 c_3 - 20 c_4 \right) H^2
+c_1 c_2^2 (c_1 + 24 c_3 + 40 c_4) m^2\,,
\\
{\cal M}_4&=&3 \left( 3 c_2 + 8 c_4 \right) \left(  3 c_1 - 12 c_3 - 4 c_4 \right)^2H^4
\cr &&
-3 c_2 \left( 3 c_1 - 12 c_3 - 4 c_4 \right) \left[
c_1 \left( 3 c_1 + 2 c_2 + 32 c_4 \right) + 4 c_2 \left( 3 c_3 + 5 c_4 \right)
\right] H^2 m^2 
+6 c_1^2 c_2^2 (c_1 + 6 c_3 + 26 c_4) m^4\,,
\\
{\cal M}_2&=&9 c_2 \left[ \left( 3 c_1 - 12 c_3 - 4 c_4 \right) H^2 -  c_1^2 m^2\right]
\left[
   \left( c_2 + 16 c_4 \right) \left( 3 c_1 - 12 c_3 - 4 c_4 \right) H^2 - c_1 c_2 \left(c_1 + 32 c_4 \right) m^2
  \right] m^2\,,
\\
{\cal M}_0
& = & 
216 \, c_2^2 \, c_4 \left[ \left( 3 c_1 - 12 c_3 - 4 c_4 \right) H^2 - c_1^2 m^2 \right]^2 m^4\,.
\eea
 As in the vector sector, we require  the kinetic  coefficient ${\cal K}$ to be strictly positive, for any value of $k$. Taking into account $c_2>0$, inspection of the expressions  (\ref{eqn:calK-def}) and (\ref{eqn:DE-def}) shows that this is the case provided the two following conditions are satisfied (except in the case $c_1=c_2$ discussed just below):
 \begin{equation}
 \label{no-ghost-scalar}
 0<c_1< 2 c_2 \,, \qquad 
 c_1^2 m^2 + \left( - 3 c_1 + 12 c_3 + 4 c_4 \right) H^2>0
 \qquad (c_1\ne c_2)\,.
 \end{equation}
 In the special case $c_1=c_2$, the kinetic coefficient ${\cal K}$ is always positive (assuming $c_2>0$) since it reduces to
 \beq
 \calK
 = \frac{m^2 k^4 c_2}{ \tilde k^2+3m^2 c_2}\qquad (c_1=c_2)\,.
\eeq

The sound speed of the scalar mode can be read from the action (\ref{action-scalar}), which gives
\begin{equation}
c_S^2 = -\frac{\cal M}{\cal K} \, \frac{a^2}{k^2} \; .
\end{equation}
The full explicit expression of $c_S^2$ is lengthy and not very illuminating, but in the high momentum limit, $k^2 / a^2 \gg m^2, H^2$,  it reduces to
\begin{equation}
\label{cs2_high_k}
c_S^2 \simeq -\frac{4 c_2 (c_3 + c_4)}{c_1(2c_2-c_1)}  \qquad ( k^2 / a^2 \gg m^2 , H^2 )
\end{equation}
Therefore, taking into account the first condition in (\ref{no-ghost-scalar}), we find that there is  no gradient instability in the scalar sector, i.e. $c_S^2 > 0$, provided
\beq
c_3+c_4<0\,.
\eeq

In the limit ${\cal H} \equiv H / M \to 0$ in the theory with dilaton-like symmetry ($b (\Phi) = e^{M \Phi}$), we have
\begin{eqnarray}
\frac{{\cal K}}{m^2 k^4} & \simeq & {\cal H} \, \tilde k^{-2} \,
\frac{\tilde c_1 \left( - \tilde c_1 + 2 \tilde c_2 \right) \tilde k^2
 + 3 \tilde c_2 \left( - 3 \tilde c_1 + 12 \tilde c_3 + 4 \tilde c_4 \right) H^2}{\tilde c_2 \tilde k^2 + 3 \left(  - 3 \tilde c_1 + 12 \tilde c_3 + 4 \tilde c_4 \right) H^2}
\nonumber\\
\frac{\cal M}{m^2 k^4} & \simeq & {\cal H} \, \frac{4 \tilde c_2^2 ( \tilde c_3 + \tilde c_4) \tilde k^4 + \tilde c_2 \left( 3 \tilde c_1 - 12 \tilde c_3 - 4 \tilde c_4 \right)  \left(2 \tilde c_1 - 5 \tilde c_2 - 12  \tilde c_3 - 20 \tilde c_4 \right) H^2 \tilde k^2 + 3 \left( 3 \tilde c_2 + 8 \tilde c_4 \right) \left(  3 \tilde c_1 - 12 \tilde c_3 - 4 \tilde c_4 \right)^2 H^4}{\left[ \tilde c_2 \tilde k^2 + 3 \left(  - 3 \tilde c_1 + 12 \tilde c_3 + 4 \tilde c_4 \right) H^2 \right]^2}
\nonumber\\
\end{eqnarray}
where $\tilde c_i \equiv c_i / \mathcal{H}$, assuming $\tilde c_i \sim {\cal O}(1)$ as discussed at the end of Subsection \ref{subsec:vector}. Thus in the limit ${\cal H} \to 0$, the scalar sector does not exhibit any pathological behaviors such as infinite or vanishing propagation speed, which would typically be an indicator of strong coupling at nonlinear level.

\subsection{Summary of stability condition}
\label{subsec:summary-stability}

In this subsection, we summarize the conditions for the perturbations to be stable and we reexpress them in term of the derivatives of the potential, via the relations (\ref{ci-explicit}). We recall that these conditions were derived by assuming a de Sitter solution in pure massive gravity, i.e. without matter. This implies in particular $\rho_g+P_g=0$, giving the  relation
\beq
\label{dS}
2\U'+ \frac{1}{\bar{N}}\left(2\E'-\bar\E\right)=0,
\eeq
which we are going to use below to express $\bar\E$ in terms of $\U'$, $\E'$ and $\bar{N}$.

As shown in the previous subsections, the tensor sector always has a healthy kinetic term and no gradient instability.  
Stability in the vector sector requires $c_2>0$, i.e.
\beq
2 {\cal U}' + \frac{{\cal E}''}{\bar{N}} > 0\,,
\eeq
and, according to (\ref{m_T}),
\beq
m_T^2=4m^2 \left(\U'+\frac{\E'}{\bar{N}}+\U_t''+\frac{\E_t''}{\bar{N}}\right)>0\,.
\eeq
In the scalar sector, we need  the two additional conditions
(\ref{no-ghost-scalar}) to avoid ghost instabilities, except if
$c_1=c_2$ i.e. ${\cal E}''=0$. These conditions translate into
\begin{equation}
2 {\cal U}' - \frac{{\cal E}''}{\bar{N}} > 0 \; , \qquad 
\frac{m^2}{H^2} {\cal U}'^2 - 2 {\cal U}' 
+ \frac{{\cal E}'}{\bar{N}} 
- 6 \left( {\cal U}_s'' + \frac{{\cal E}_s''}{\bar{N}} \right) 
- 2 \left( {\cal U}_t'' + \frac{{\cal E}_t''}{\bar{N}} \right) > 0
\qquad ({\cal E}''\ne 0) \;.
\label{eqn:noghost-scalar}
\end{equation}
Finally, the high momentum sound speed (\ref{cs2_high_k}) is given by
\begin{equation}
c_S^2 \simeq 2 \, \frac{{\cal U}' + {\cal E}' / \bar{N} + 2 \left( {\cal U}_s'' + {\cal E}_s'' / \bar{N} + {\cal U}_t'' + {\cal E}_t'' / \bar{N} \right)}{2 {\cal U}' - {\cal E}'' / \bar{N}} \qquad ( k^2 / a^2 \gg m^2 , H^2 )
\end{equation}
and therefore,  the absence of gradient instability is guaranteed by the extra condition 
\begin{equation}
{\cal U}' + \frac{{\cal E}'}{\bar{N}} + 2 \left( {\cal U}_s'' + \frac{{\cal E}_s''}{\bar{N}} + {\cal U}_t'' + \frac{{\cal E}_t''}{\bar{N}} \right) > 0 \;.
\label{eqn:nogradient-scalar}
\end{equation}

Putting everything together,  for ${\cal E}''\ne 0$, the following
conditions must be satisfied simultaneously: 
\begin{eqnarray}
&& 2\U' > \Big\vert \frac{\E''}{\bar{N}} \Big\vert \; , \\
&& \U'+\frac{\E'}{\bar{N}}+\U_t''+\frac{\E_t''}{\bar{N}} > 0 \; , \\
&& {\cal U}' + \frac{{\cal E}'}{\bar{N}} + 2 \left( {\cal U}_s'' + \frac{{\cal E}_s''}{\bar{N}} + {\cal U}_t'' + \frac{{\cal E}_t''}{\bar{N}} \right) > 0 \; , \\
&& \frac{m^2}{H^2} {\cal U}'^2 - 2 {\cal U}' + \frac{{\cal E}'}{\bar{N}} - 6 \left( {\cal U}_s'' + \frac{{\cal E}_s''}{\bar{N}} \right) - 2 \left( {\cal U}_t'' + \frac{{\cal E}_t''}{\bar{N}} \right) > 0 \; .
\end{eqnarray}
Let us mention the special case  $\E''=0$, which corresponds to $c_1=c_2$. In this case, the first condition above reduces to $\U'>0$ and the last condition is no longer required.  

Since generically the number of arbitrary functions exceeds the number of the constraint inequalities, it is naturally expected that there is a broad region in parameter space where the system is free from instabilities. In the next section, we will give an explicit example, which shows the existence of such  stable solutions.
When matter is present in the universe, the instability conditions are more involved, but matter will eventually be diluted and become negligible, so that  the conditions derived  above remain useful for the long-term stability.

\section{Explicit example}
\label{sec:example}

In this section we consider a simple explicit example, especially
focusing on the behavior of the effective gravitational field which is
induced by the graviton mass term. We shall see that there actually
exists a healthy cosmological solution satisfying all conditions for
healthy perturbations derived in the previous section. 

\subsection{Ansatz for the potential}

In the simple example that we investigate in this section, the explicit
forms of $\calU$ and $\calE$ are assumed to be 
 \begin{align}
 \calU (\calK^{IJ} \,, f_{IJ}) 
 &= u_0 + u_1 \calK^{IJ} f_{IJ}
 + \frac{1}{2} \Bigl[ u_{2s} ( \calK^{IJ} f_{IJ})^2
 + u_{2q} \calK^{IJ} f_{J K} \calK^{KL} f_{LI} \Bigr] \,, \\
 \calE (\Gamma^{IJ} \,, \xi^I \,, f_{IJ}) 
 &= v_0 + v_1 \Gamma^{IJ} f_{IJ}
 + \frac{1}{2} \Bigl[ v_{2s} (\Gamma^{IJ} f_{IJ})^2
 + v_{2q} \Gamma^{IJ} f_{J K} \Gamma^{KL} f_{LI} \Bigr]
 + \frac{1}{2} w_2 \xi^I \xi^J f_{IJ} \,,
  \label{eqn:UandEexample}
 \end{align}
 where the coefficients $u_0$, $u_1$, etc. are constants.

 In terms of the ratio $X\equiv b/a$, introduced in (\ref{X}), 
 we can write 
the background values of ${\cal U}$ and ${\cal E}$ as 
\begin{equation}
\bar{{\cal U}} = u_0 + 3 \, u_1 \, X^2 + \frac{3}{2} \, u_2 \, X^4\,, \quad
\bar{{\cal E}} = v_0 + 3 \, v_1 \, X^2 + \frac{3}{2} \, v_2 \, X^4\,,
\end{equation} 
with
\begin{equation}
 u_2 \equiv 3 u_{2 s} + u_{2 q}\,, \quad
 v_2 \equiv 3 v_{2 s} + v_{2 q}\,.
\end{equation}
Following the definitions (\ref{V'}) and  (\ref{V''1}-\ref{V''2})  of
the various parameters derived from the potential, we obtain explicitly 
\begin{eqnarray}
 {\cal U}' & = & u_1 \, X^2 + u_2 \, X^4 \; , \quad
 {\cal U}_s'' = u_{2s} \, X^4 \; , \quad 
 {\cal U}_t'' = u_{2q} \, X^4 \; , \nonumber\\
 {\cal E}' & = & v_1 \, X^2 + v_2 \, X^4 \; , \quad
 {\cal E}_s'' = v_{2s} \, X^4 \; , \quad 
 {\cal E}_t'' = v_{2q} \, X^4 \; , \quad
 {\cal E}'' = w_2 \, X^2\,. \label{eqn:potentials-example}
\end{eqnarray}

\subsection{Background equations}

Following (\ref{Friedman-eq}), the  background equations are given by
\begin{align}
 3 M_{\rm Pl}^2H^2 = \rho_g + \rho_m \,, \qquad
 M_{\rm Pl}^2\left(2 \frac{\dot{H}}{\bar{N}} + 3 H^2 \right)
 = - (P_g + P_m)\,,
\end{align}
where $\rho_m$ and $P_m$ represent the energy density and pressure
from usual matter and $\rho_g$ and $P_g$ are those from the effective
gravitational field. In the current model, their explicit forms are given as
\begin{align}
 \rho_g &= M_{\rm Pl}^2m^2 
 \left( u_0 + 3 u_1 X^2 + \frac{3}{2} u_2 X^4 \right) \,,
\end{align}
 and
\begin{align}
 P_g  &= M_{\rm Pl}^2m^2 
 \left[ - u_0 - u_1 X^2 + \frac{1}{2} u_2 X^4 
 +\frac{1}{\bar{N}}
 \left( -v_0 - v_1 X^2 + \frac{1}{2} v_2 X^4 \right)\right] \,.
\end{align}

\subsection{De Sitter solution}

In the absence of matter ($\rho_m=P_m=0$) and assuming the dilaton-like
symmetry ($\dot{b}/b=M$), $X$ and $H$ are determined by the algebraic
equations (\ref{eqn:eq-X-vacuum}) and (\ref{eqn:H-vacuum}). One can
instead solve these equations with respect to two parameters of the
model, e.g. $u_0$ and $v_0$, as
\begin{equation}
 u_0 = \frac{3H^2}{m^2} - 3u_1X^2 - \frac{3}{2}u_2X^4\,, \quad
  v_0 = 2 \frac{M}{H} \left(u_1X^2+u_2X^4 \right) - v_1X^2 + \frac{1}{2} v_2X^4 \,.
\end{equation} 
One can also solve (\ref{eqn:potentials-example}) with respect to
($u_1$, $u_{2s}$, $u_{2q}$, $v_1$, $v_{2s}$, $v_{2q}$, $w_2$) to express
them in terms of (${\cal U}'$, ${\cal U}_s''$, ${\cal U}_t''$, ${\cal
E}'$, ${\cal E}_s''$, ${\cal E}_t''$, ${\cal E}''$) and $X$. Hereafter
we thus consider ($H$, $X$, ${\cal U}'$, ${\cal U}_s''$, ${\cal U}_t''$,
${\cal E}'$, ${\cal E}_s''$, ${\cal E}_t''$, ${\cal E}''$) as
independent parameters. Since $X$ is constant, $\bar{N}$ is determined
from (\ref{Udot2Up}) (with $\dot{\bar{{\cal U}}}=0$ and $\dot{b}/b=M$)
as 
\begin{equation}
 \bar{N} = \frac{M}{H}.
\end{equation}

With $\dot{b}/b=M$, the stability condition summarized in
subsection~\ref{subsec:summary-stability} become
\begin{eqnarray}
&& 2\U' > \vert \E'' \vert \, \frac{H}{M} \; , \\
&& \U'+\U_t''+ \left( \E' + \E_t'' \right) \frac{H}{M}> 0 \; , \\
&& {\cal U}' + 2 {\cal U}_s'' + 2 {\cal U}_t'' + \left( {\cal E}' + 2 {\cal E}_s'' + 2 {\cal E}_t'' \right) \frac{H}{M} > 0 \; , \\
&& \frac{m^2}{H^2} {\cal U}'^2 - 2 \left( {\cal U}' + 3 {\cal U}_s'' + {\cal U}_t'' \right) + \left( {\cal E}' - 6 {\cal E}_s'' - 2 {\cal E}_t'' \right) \frac{H}{M} > 0 \; ,
\end{eqnarray}
assuming $M > 0$. As there are $9$ independent parameters in the $4$
constraint inequalities, none of which is conflicting with each other,
this suffices to show the stability of the present example. This
confirms  the existence of stable, pure de Sitter solutions in this
theory.

\section{Comparison with previous results}
\label{sec:comparison}

\subsection{dRGT}

As briefly reviewed in the introduction, the dRGT theory with Minkowski
fiducial metric allows for self-accelerating open FLRW
solutions~\cite{Gumrukcuoglu:2011ew}. If we replace the fiducial metric
with a de Sitter or FLRW one, then the theory permits self-accelerating
flat/closed/open FLRW solutions. However, for those solutions the
quadratic action for perturbations does not contain kinetic terms for
three among five degrees of freedom~\cite{Gumrukcuoglu:2011zh}. This is
the origin of the nonlinear ghost instability found in
\cite{DeFelice:2012mx}. Based on this nonlinear instability and the
linear instability called Higuchi ghost~\cite{Higuchi:1986py}, it was
argued that all FLRW solutions in the dRGT theory are unstable.

The analysis in Section~\ref{sec:perturbations} can reproduce the key
result of \cite{Gumrukcuoglu:2011zh}, namely the  vanishing of kinetic terms for
three among the five degrees of freedom for the perturbations. The quadratic
action for the perturbations around self-accelerating solutions in the dRGT
theory has a peculiar structure. As shown in an appendix of
\cite{Gumrukcuoglu:2011zh}, it is in the general form
(\ref{eqn:Smg-general}) but with
\begin{equation}
 c_0=c_1=c_2=0\,,\qquad c_3+c_4=0\,,\qquad
  (\mbox{dRGT self-accelerating branch})\,.\label{eqn:dRGT}
\end{equation} 
Among the properties (\ref{eqn:dRGT}), $c_0=0$ always holds in theories 
without BD ghost. On the other hand, $c_1=c_2=0$ is the origin of the
vanishing kinetic terms. Indeed, $c_2=0$ already implies that $\mu^2$
defined in (\ref{eqn:mu-def}) vanishes and thus the kinetic term for
$E_i$ in the vector action (\ref{action-vector}) vanishes. The
coefficient ${\cal K}$ defined in (\ref{eqn:calK-def}) also vanishes
when $c_1=c_2=0$, meaning that the kinetic term of $E$ in the scalar
action (\ref{action-scalar}) vanishes.

\subsection{Comelli et al}

In \cite{Comelli:2013tja}, the fiducial metric was supposed to be
strictly euclidean, i.e. without the scale factor $b$ that we have
introduced in (\ref{eqn:3dfiducialmetric}). Their results can thus be
reproduced by simply setting $\dot{b}/b=0$ in our equations. 
Indeed, in this case, 
(\ref{eqn:constraint}) implies the  constraint 
 \beq
 H(2{\cal E}'-\bar{{\cal E}})=0\,.
 \eeq
For a generic choice of the function ${\cal E}$, if $H\ne 0$ then this
equation would fix $X=b/a$ to a constant, which is a root of 
$2{\cal E}'-\bar{{\cal E}}$. However, since $b$ is constant here,
$X={\rm const.}$ would mean $H=0$. To get a non-trivial cosmology, the function
${\cal E}$ is thus severely restricted (so that 
$2{\cal E}'-\bar{{\cal E}}$ vanishes identically) and the effective
gravitational energy density and pressure reduce to 
\beq
 \rho_g\equiv M_{\rm Pl}^2m^2 \bar{{\cal U}}, \qquad  
 P_g\equiv M_{\rm Pl}^2m^2
 \left(2\U'-\bar{{\cal U}}\right)\,.
\eeq

In this situation, an equation of state $P_g=-\rho_g$, corresponding to
a de Sitter solution, necessarily requires 
  \begin{equation}
 \U' = 0 \qquad  (\mbox{de Sitter} )\,.
\end{equation} 
This implies that the coefficients $c_1$ and $c_2$ defined in (\ref{ci-explicit}) 
vanish. As explained in the previous subsections, this inevitably means
that the vector and scalar modes have vanishing kinetic terms, in agreement with  the results of \cite{Comelli:2013tja}. 
In that work, in order to avoid this pathological behaviour, they introduce a slight deviation from the equation of state $P_g=-\rho_g$. By contrast, in our case, the extra freedom due to the scale factor $b$ enables us to find strictly de Sitter solutions that are not pathological.

\section{Summary and discussions}

In this work,  we have investigated FLRW cosmological solutions and
their stability in a class of massive gravity theories with five degrees
of freedom. The theories we have considered for this study are those
that respect the $3$-dimensional maximal symmetry in the space of
St\"{u}ckelberg fields, but not necessarily the $4$-dimensional one, in
accord with the symmetry of FLRW
cosmology~\cite{Comelli:2013paa,Comelli:2013txa}. In any theory of
massive gravity written in the unitary gauge, diffeomorphism invariance
is broken by the graviton mass term at cosmological scales. It is thus 
rather natural to suppose that the low energy effective field theory of
massive gravity respects only the symmetry of the cosmological background,
i.e. the $3$-dimensional maximal symmetry.

After a detailed study of linear perturbations, we have shown that,
unlike in the dRGT theory, all five degrees of freedom around
self-accelerating FLRW backgrounds generically have non-vanishing
kinetic terms in this theory. This means that  the setup developed in the present paper evades the no-go result for massive gravity cosmology found recently in
\cite{DeFelice:2012mx}. 

It is important to stress that our model differs not only from the dRGT
theory but also from  that studied in a previous work on the cosmology
of rotation-invariant massive gravity~\cite{Comelli:2013tja}. In that
previous study, it was found that the kinetic terms of three among five
degrees of freedom vanish on de Sitter backgrounds. By contrast, in the
model considered  in the present paper, all five degrees of freedom
generically have non-vanishing kinetic terms on de Sitter as well as
non-de Sitter FLRW backgrounds. The main difference between our
theoretical setup and the previous one in \cite{Comelli:2013tja} is that
we allow the graviton mass term to depend on the temporal
St\"{u}ckelberg field in a non-trivial way. In the dRGT limit, such a
non-trivial dependence manifests itself as a de Sitter or FLRW fiducial
metric. (Note however that three among the  five degrees of freedom have
vanishing kinetic terms in any self-accelerating FLRW backgrounds in
dRGT with any FLRW fiducial metric.) 
For illustration, we have also
presented simple models that  allow for stable de Sitter
solutions. Although  we have restricted our analysis, for simplicity, to
cases assuming an additional dilaton-like symmetry in the
St\"{u}ckelberg field space, our results correspond to a generic feature
of rotation-invariant  massive gravity theory with a non-trivial
fiducial metric.

\begin{acknowledgments}
 We would like to thank Luigi Pilo for very interesting discussions, as
 well as the Yukawa Institute for Theoretical Physics at Kyoto
 University, where part of this work was done during
 the YITP Workshop 
 YITP-X-13-03 on
``APC-YITP collaboration: Mini-Workshop on Gravitation and Cosmology''.
 SM thanks APC (CNRS-Universit\'e Paris 7)
 for hospitality during the beginning of this work. This work was
 supported by the WPI Initiative, MEXT, Japan. 
 DL was supported in part by the  ANR (Agence Nationale de la Recherche)
 grant STR-COSMO ANR-09-BLAN- 0157-01. 
 The work of SM was supported by Grant-in-Aid for Scientific Research
 24540256, 21111006 and 21244033, MEXT, Japan. 
 A.N. was supported in part by JSPS Postdoctoral Fellowships for
 Research Abroad and Grant-in-Aid for JSPS Fellows No. 26-3409.
\end{acknowledgments}

\appendix

\section{Quadratic expansion of Einstein-Hilbert action}
\label{app:EHaction}

In this appendix, we give some details about the expansion of the
Einstein-Hilbert action up to quadratic order in linear perturbations
around a spatially flat FLRW background solution. 

We employ the ADM formalism in which the metric is given by
(\ref{ADM-line}). The Einstein-Hilbert action can then be written in the
form
\beq
S_{\rm EH} = \frac{M_{\rm Pl}^2}{2} \int \dd^4x \, R = \frac{M_{\rm Pl}^2}{2} \int \dd^4 x \, N
 \sqrt{\gamma} \Bigl( K_{ij} K^{ij}-K^2+{}^{(3)}R \Bigr) \,,
\label{action-EH}
\eeq
where $\gamma=$ det$(\gamma_{ij})$, ${}^{(3)} R$ is the Ricci scalar
associated with the 3-D spatial metric $\gamma_{ij}$, and 
\beq
K_{ij}=\frac{1}{2N}\left( \dot{\gamma}_{ij}-N_{i|j}-N_{j|i}\right)
\eeq
is the extrinsic curvature tensor (the symbol $|$ denotes the spatial
covariant derivative associated with the spatial metric
$\gamma_{ij}$). Here the indices on $K_{ij}$ are raised (lowered) by
$\gamma^{ij}$ ($\gamma_{ij}$), and $K \equiv K^i_{\, i}$.

We now expand the metric around the flat FLRW background. The metric
perturbations are described by the lapse perturbation $\delta N$, the
shift $N^i$ and the perturbations $h_{ij}$ of the
three-dimensional metric, as described at the beginning of
Section~\ref{sec:perturbations}. 
To compute the Einstein-Hilbert action up to quadratic order, one needs
to compute $K^i_{\, j}$ up to linear order and ${}^{(3)}R$ up to
quadratic order, as is clear from (\ref{action-EH}). One can write, up
to quadratic order, 
\beq
K_{ij} K^{ij}-K^2 = 6H^2-4H K +\delta K^i_{\, j}\delta K^j_{\, i}-\delta K^2 \; .
\eeq
Moreover, by integration by parts, one has
\beq
\int \dd^4x \sqrt{-g} \, H K
 = \int \dd^4x \sqrt{-g} \,  H \, \nabla_\mu n^\mu
 \rightarrow - \int \dd^4x \sqrt{-g} \, \frac{\dot H}{N}\,,
\eeq
where $n^\mu$ is the unit vector normal to the constant time hypersurfaces.
Therefore, the Einstein-Hilbert action can be rewritten, up to total derivatives, as
\bea
\!\!\!\!\!\! && S_{\rm EH} = M_{\rm Pl}^2 \int \dd^4x \sqrt{-g} \, (3H^2)+\tilde S_{EH} \nonumber\\
\!\!\!\!\!\! && \quad  \tilde S_{\rm EH} \equiv 2 M_{\rm Pl}^2 \int \dd^4 x \sqrt{\gamma} \, \dot H + \frac{M_{\rm Pl}^2}{2} \int \dd^4x N\sqrt{\gamma}
\Bigl( \delta K^i_{\, j}\delta K^j_{\, i}-\delta K^2+{}^{(3)}R \Bigr) \,,
\label{tS_EH-appendix}
\eea
where the explicit expressions are given by
\bea
\delta K^i_{\, j} \!\!\! & = & \!\!\! \frac{1}{2\bar{N}}
\left(
 \dot h^i_{\, j}-2H\d N\d^i_j-\delta^{ik}\delta_{jl}\, \partial_k N^l-\partial_j N^i\right) \nonumber\\
\delta K \!\!\! & = & \!\!\! \frac{1}{2\bar{N}}
 \left(\dot h-3H\d N -2\partial_k N^k\right) \nonumber\\
{}^{(3)}R \!\!\! &=& \!\!\! \frac{1}{a^2} \bigg( -\p^2 h+\p_i\p_j h^{ij} 
+h^{ij}\, \p^2 h_{ij}+h^{ij}\, \p_i\p_j h -2\, h^{ij} \, \p_i \p^k h_{kj}+\frac34 \, \p_i h_{jk} \, \p^ih^{jk}
\cr
&& \qquad
-\frac12 \p_k h_{ij}\, \p^j h^{ik}-\p_i h^{ij}\, \p^kh_{kj}-\frac14 \p^kh \, \p_k h+\p_i h^{ij}\, \p_j h \bigg) \, ,
\eea
In the above expressions and in the following,  the spatial indices are
raised and lowered by $\delta^{ij}$ and $\delta_{ij}$,
respectively. Substituting the above expressions into
(\ref{tS_EH-appendix}) and using the following formula valid up to the
quadratic order, 
\beq
\sqrt{\gamma}=a^3\left(1+\frac12 h -\frac14 h_{ij}h^{ij}+\frac18 h^2\right)\,,
\eeq
one finally obtains the quadratic action for the Einstein-Hilbert term
in the form
\bea
\tilde S_{\rm EH}^{(2)} \!\!\! & = & \!\!\! M_{\rm Pl}^2 \int \dd^4x \, \bar{N} a^3\Bigg\{\frac{1}{8\bar{N}^2} \dot h_{i j}  \dot h^{ij}-\frac{1}{8\bar{N}^2} \dot h^2+ \frac{H}{\bar{N}^2}\d N \dot h
-\frac{1}{2\bar{N}} \dot h_{ij}\p^iN^j+\frac{1}{2\bar{N}} \dot h\p_k N^k
-3 H^2\frac{\d N^2}{\bar{N}^2}
\cr
&& \!\!\!
+\frac{1}{4\bar{N}^2}\p_i N_j \left(\p^iN^j+\p^j N^i\right)
-\frac{2H}{\bar{N}^2}\d N \p_i N^i-\frac{1}{2\bar{N}^2} (\p_k N^k)^2
+\frac{1}{a^2}\bigg[\left(\frac{\d N}{\bar{N}}+\frac12 h\right)\left(-\p^2 h+\p_i\p_j h^{ij} \right)
\cr
&& \!\!\!
+h^{ij}\, \p^2 h_{ij}+h^{ij}\, \p_i\p_j h 
-2\, h^{ij} \, \p_i \p^k h_{kj}+\frac34 \, \p_i h_{jk} \, \p^ih^{jk}
-\frac12 \p_k h_{ij}\, \p^j h^{ik}-\p_i h^{ij}\, \p^kh_{kj}
-\frac14 \p^kh \, \p_k h+\p_i h^{ij}\, \p_j h
\bigg]
\cr
&& \!\!\!
+2\frac{\dot H}{\bar{N}}\left(\frac18 h^2-\frac14 h_{ij}h^{ij}\right)
\Bigg\} \, .
\label{action_EH_quad}
\eea
We now decompose this action into its tensor, vector and scalar components.

\subsubsection{Tensor sector}

The lapse and shift are unperturbed and $h_{ij}$ is transverse and traceless:
\beq
\p^ih_{ij}=0,\qquad h=0\,.
\eeq
The quadratic part of the Einstein-Hilbert action for tensor modes is
\bea
\tilde S_{\rm EH}^{(2)T}&=&\int \dd^4x \,\bar{N} a^3\left( \frac{1}{8\bar{N}^2} \dot h_{i j}  \dot h^{ij}-\frac{1}{8a^2} \, \p_k h_{ij} \, \p^k h^{ij}-\frac{\dot H}{2}h_{ij}h^{ij}
\right) \,.
\eea

\subsubsection{Vector sector}

The only perturbations are $N^i$ and
\beq
h_{ij}=\p_i E_j+\p_j E_i,
\eeq
with 
\beq
\p_i N^i=0\,, \qquad \p_i E^i=0\,.
\eeq
Substituting these into the action (\ref{action_EH_quad}), and
simplifying via integrations by parts, one finally obtains
\beq
\tilde S_{\rm EH}^{(2)V} = 
\int \dd^4x \, \frac{a^3}{\bar{N}}
\left( \frac14 \p_i \dot E_j  \p^j\dot E^{j}+\frac{1}{4} \p_i N_j \, \p^i N^j-\frac12 \p_i\dot E_j \p^i N^j
-\bar{N}\dot H \, \p_iE_j \p^i E^j
\right) \,.
\eeq

\subsubsection{Scalar sector}

The metric perturbations are described by $\delta N$, and
\beq
N^i=\p^i B, \qquad h_{ij}=2C \d_{ij}+2\p_i\p_j E\,.
\eeq
After a lot of simplifications, one finds
\bea
\tilde S_{\rm EH}^{(2)S}&=&\int \dd^4x \, a^3\bigg\{\frac32 \dot C^2+\dot C\p^2 \dot E +\frac12 (\p^2\dot E)^2+\frac32 H^2\d N^2
-3H \d N \dot C -H\d N \p^2\dot E +H \d N \p^2 B
+\frac12 (\p^2 B)^2
\cr 
&& 
-\dot C\p^2 B -\p^2\dot E \p^2 B
 +\frac{1}{2a^2} \Bigl[ -4\p^2C \d N+2(\p C)^2 \Bigr]
+\dot H \Bigl[ 3C^2+2C \p^2 E -(\p^2 E)^2 \Bigr]
\bigg\}\,.
\eea

\section{Condition for no BD ghost at quadratic order} 
\label{app:noBD}

In this appendix, we extend our  analysis of the linear perturbations to
include the case of an arbitrary coefficient $c_0$ in
(\ref{eqn:Smg-general}). We then show that a non-zero $c_0$ leads to the
presence of a sixth propagating degree of freedom, which is known to
lead to a BD ghost at the nonlinear level. The fact that the potential
(\ref{eqn:potential-general}) leads to $c_0=0$ is thus consistent with
the property that this potential is  free of the BD-ghost at all orders
\cite{Comelli:2013paa,Comelli:2013txa}. Here we restrict our
consideration to the flat FLRW background without matter, as the
inclusion of matter leads to complicated expressions, making our
analysis less transparent.\footnote{
We have verified that the condition for the absence of the BD ghost
($c_0=0$) is unchanged by including matter. 
}

We thus start from  (\ref{action-tot}) with (\ref{action_EH_quad}) and
(\ref{eqn:Smg-general}),  without any restriction on the coefficients
$c_i$. They can also depend on time, in contrast with our assumption in Section \ref{sec:perturbations}. 
We then decompose the perturbations into tensor, vector and scalar modes as
in (\ref{tensor-dec}), (\ref{vector-dec}) and (\ref{scalar-dec}),
respectively. The actions for the tensor and vector modes are identical
to those derived in Subsections \ref{subsec:tensor} and
\ref{subsec:vector} and the stability conditions in these sectors are thus unchanged.

By contrast, the action in the scalar sector is different in general. After
integrating out the non-dynamical lapse $\delta N$ and shift component $B$, the
scalar quadratic action  contains two degrees of freedom and
can be written in the  following matrix form 
\begin{equation}
S_S^{(2)} = \int \dd t \, \dd^3k \, a^3 \left[ \dot{\delta}^\dagger \, \tilde{\cal T} \, \dot{\delta} + \left( \delta^\dagger \, \tilde{\cal X} \, \dot{\delta} + {\rm h.c.} \right) + \delta^\dagger \, \tilde{\Omega}^2 \, \delta \right]
\label{action-sca-app}
\end{equation}
where $\delta \equiv \left( C , E \right)^T$,  the first two coefficient
matrices are given by 
\begin{eqnarray}
\tilde{T} \!\!\! & = & \!\!\! \frac{m^2}{\left( \frac{k^2}{a^2} + 3 \, m^2 c_2 \right) H^2 - m^4 \, c_0 \, c_2} \left[
\begin{array}{cc}
\left( \frac{k^2}{a^2} + 3 \, m^2 c_2 \right) c_0 & - m^2 k^2 c_0 \, c_2 \\
- m^2 k^2 c_0 \, c_2 & k^4 H^2 c_2
\end{array}
\right] 
\label{Ttil-mat}\\
\tilde{X} \!\!\! & = & \!\!\! \left[
\begin{array}{cc}
0 & m^2 k^2 H \, \frac{\frac{k^2}{a^2} \left( c_1 - c_2 \right)}{\left( \frac{k^2}{a^2} + 3 \, m^2 c_2 \right) H^2 - m^4 \, c_0 \, c_2} \\
0 & 0
\end{array}
\right]
\label{Xtil-mat}
\end{eqnarray}
and $\tilde{\Omega}^2$ is a $2\times2$ Hermitian matrix, whose rather
lengthy expression is not needed for our current purpose.  

In general, the above action contains two dynamical degrees of freedom.
It is known that one of these scalar modes leads to ghost-like
instabilities at the nonlinear level. To get rid of this pathological
mode, one must ensure that only one dynamical degree of freedom appears
in the scalar action. A necessary condition is that  the determinant of
the matrix $\tilde{T}$, given in (\ref{Ttil-mat}), vanishes, namely 
\begin{equation}
\det \tilde{T} = m^4 k^4 \, \frac{c_0 \, c_2}{\left( \frac{k^2}{a^2} + 3 \, m^2 c_2 \right) H^2 - m^4 \, c_0 \, c_2} =0\,.
\end{equation}
Since stability in the vector sector  requires $c_2>0$, this imposes the condition 
\begin{equation}
c_0 = 0 \; .
\end{equation}
Once this condition is imposed, one can check that  the scalar  mode $C$
becomes non-dynamical. After integrating it out, one recovers the scalar
quadratic action (\ref{action-scalar}) with only one dynamical degree of
freedom, $E$. 

In view of the above discussion, it appears natural that the theory
considered in the main text leads to $c_0 = 0$, as explicitly shown in
(\ref{ci-explicit}). This is a non-trivial consistency check for the
theory, which should be free of the BD ghost at all orders
\cite{Comelli:2013paa,Comelli:2013txa}.



\begin{thebibliography}{99}

\bibitem{Fierz:1939ix} 
  M.~Fierz and W.~Pauli,
  Proc.\ Roy.\ Soc.\ Lond.\ A {\bf 173}, 211 (1939).

\bibitem{vanDam:1970vg} 
  H.~van Dam and M.~J.~G.~Veltman,
  Nucl.\ Phys.\ B {\bf 22}, 397 (1970).

\bibitem{Zakharov:1970cc} 
  V.~I.~Zakharov,
  JETP Lett.\  {\bf 12}, 312 (1970)
  [Pisma Zh.\ Eksp.\ Teor.\ Fiz.\  {\bf 12}, 447 (1970)].

\bibitem{Vainshtein:1972sx} 
  A.~I.~Vainshtein,
  Phys.\ Lett.\ B {\bf 39}, 393 (1972).

\bibitem{Boulware:1973my} 
  D.~G.~Boulware and S.~Deser,
  Phys.\ Rev.\ D {\bf 6}, 3368 (1972).
  
\bibitem{Hinterbichler:2011tt} 
  K.~Hinterbichler,
  Rev.\ Mod.\ Phys.\  {\bf 84}, 671 (2012)
  [arXiv:1105.3735 [hep-th]].

\bibitem{deRham:2014zqa} 
  C.~de Rham,
  arXiv:1401.4173 [hep-th].

\bibitem{deRham:2010ik} 
  C.~de Rham and G.~Gabadadze,
  Phys.\ Rev.\ D {\bf 82}, 044020 (2010)
  [arXiv:1007.0443 [hep-th]].

\bibitem{deRham:2010kj} 
  C.~de Rham, G.~Gabadadze and A.~J.~Tolley,
  Phys.\ Rev.\ Lett.\  {\bf 106}, 231101 (2011)
  [arXiv:1011.1232 [hep-th]].

\bibitem{Hassan:2011hr}
  S.~F.~Hassan and R.~A.~Rosen,
  Phys.\ Rev.\ Lett.\  {\bf 108} (2012) 041101
  [arXiv:1106.3344 [hep-th]].

\bibitem{Hassan:2011ea}
  S.~F.~Hassan and R.~A.~Rosen,
  JHEP {\bf 1204} (2012) 123
  [arXiv:1111.2070 [hep-th]].

\bibitem{Kugo:2014hja} 
  T.~Kugo and N.~Ohta,
  PTEP {\bf 2014}, no. 4, 043B04
  [arXiv:1401.3873 [hep-th]].


\bibitem{D'Amico:2011jj} 
  G.~D'Amico, C.~de Rham, S.~Dubovsky, G.~Gabadadze, D.~Pirtskhalava and A.~J.~Tolley,
  Phys.\ Rev.\ D {\bf 84}, 124046 (2011)
  [arXiv:1108.5231 [hep-th]].

\bibitem{Gumrukcuoglu:2011ew} 
  A.~E.~Gumrukcuoglu, C.~Lin and S.~Mukohyama,
  JCAP {\bf 1111}, 030 (2011)
  [arXiv:1109.3845 [hep-th]].

\bibitem{Gumrukcuoglu:2011zh} 
  A.~E.~Gumrukcuoglu, C.~Lin and S.~Mukohyama,
  JCAP {\bf 1203}, 006 (2012)
  [arXiv:1111.4107 [hep-th]].

\bibitem{Hassan:2011vm} 
  S.~F.~Hassan and R.~A.~Rosen,
  JHEP {\bf 1107}, 009 (2011)
  [arXiv:1103.6055 [hep-th]].

\bibitem{Fasiello:2012rw}
  M.~Fasiello and A.~J.~Tolley,
  JCAP {\bf 1211} (2012) 035
  [arXiv:1206.3852 [hep-th]].

\bibitem{Langlois:2012hk} 
  D.~Langlois and A.~Naruko,
  Class.\ Quant.\ Grav.\  {\bf 29}, 202001 (2012)
  [arXiv:1206.6810 [hep-th]].

\bibitem{DeFelice:2012mx}
  A.~De Felice, A.~E.~Gumrukcuoglu and S.~Mukohyama,
  Phys.\ Rev.\ Lett.\  {\bf 109} (2012) 171101
  [arXiv:1206.2080 [hep-th]].

\bibitem{Higuchi:1986py} 
  A.~Higuchi,
  Nucl.\ Phys.\ B {\bf 282}, 397 (1987).

\bibitem{Chamseddine:2011bu} 
  A.~H.~Chamseddine and M.~S.~Volkov,
  Phys.\ Lett.\ B {\bf 704}, 652 (2011)
  [arXiv:1107.5504 [hep-th]].

\bibitem{Koyama:2011xz} 
  K.~Koyama, G.~Niz and G.~Tasinato,
  Phys.\ Rev.\ Lett.\  {\bf 107}, 131101 (2011)
  [arXiv:1103.4708 [hep-th]].

\bibitem{Koyama:2011yg} 
  K.~Koyama, G.~Niz and G.~Tasinato,
  Phys.\ Rev.\ D {\bf 84}, 064033 (2011)
  [arXiv:1104.2143 [hep-th]].

\bibitem{Gratia:2012wt} 
  P.~Gratia, W.~Hu and M.~Wyman,
  Phys.\ Rev.\ D {\bf 86}, 061504 (2012)
  [arXiv:1205.4241 [hep-th]];

\bibitem{Kobayashi:2012fz} 
  T.~Kobayashi, M.~Siino, M.~Yamaguchi and D.~Yoshida,
  Phys.\ Rev.\ D {\bf 86}, 061505 (2012)
  [arXiv:1205.4938 [hep-th]].

\bibitem{Volkov:2012cf} 
  M.~S.~Volkov,
  Phys.\ Rev.\ D {\bf 86}, 061502 (2012)
  [arXiv:1205.5713 [hep-th]].
 
\bibitem{Gumrukcuoglu:2012aa} 
  A.~E.~Gumrukcuoglu, C.~Lin and S.~Mukohyama,
  Phys.\ Lett.\ B {\bf 717}, 295 (2012)
  [arXiv:1206.2723 [hep-th]].

\bibitem{DeFelice:2013awa}
  A.~De Felice, A.~E.~Gumrukcuoglu, C.~Lin and S.~Mukohyama,
  JCAP {\bf 1305}, 035 (2013)
  [arXiv:1303.4154 [hep-th]].

\bibitem{D'Amico:2012zv}
  G.~D'Amico, G.~Gabadadze, L.~Hui and D.~Pirtskhalava,
  Phys.\ Rev.\ D {\bf 87}, no. 6, 064037 (2013)
  [arXiv:1206.4253 [hep-th]].

\bibitem{Huang:2012pe}
  Q.~-G.~Huang, Y.~-S.~Piao and S.~-Y.~Zhou,
  Phys.\ Rev.\ D {\bf 86} (2012) 124014
  [arXiv:1206.5678 [hep-th]].

\bibitem{Gumrukcuoglu:2013nza} 
  A.~E.~Gumrukcuoglu, K.~Hinterbichler, C.~Lin, S.~Mukohyama and M.~Trodden,
  Phys.\ Rev.\ D {\bf 88}, 024023 (2013)
  [arXiv:1304.0449 [hep-th]].

\bibitem{D'Amico:2013kya}
  G.~D'Amico, G.~Gabadadze, L.~Hui and D.~Pirtskhalava,
  Class.\ Quant.\ Grav.\  {\bf 30} (2013) 184005
  [arXiv:1304.0723 [hep-th]].

\bibitem{Gabadadze:2014kaa} 
  G.~Gabadadze, R.~Kimura and D.~Pirtskhalava,
  Phys.\ Rev.\ D {\bf 90}, 024029 (2014)
  [arXiv:1401.5403 [hep-th]].

\bibitem{DeFelice:2013tsa} 
  A.~De Felice and S.~Mukohyama,
  Phys.\ Lett.\ B {\bf 728}, 622 (2014)
  [arXiv:1306.5502 [hep-th]].

\bibitem{DeFelice:2013dua} 
  A.~De Felice, A.~E.~Gumrukcuoglu and S.~Mukohyama,
  Phys.\ Rev.\ D {\bf 88}, 124006 (2013)
  [arXiv:1309.3162 [hep-th]].

\bibitem{Gabadadze:2012tr} 
  G.~Gabadadze, K.~Hinterbichler, J.~Khoury, D.~Pirtskhalava and M.~Trodden,
  Phys.\ Rev.\ D {\bf 86}, 124004 (2012)
  [arXiv:1208.5773 [hep-th]].

\bibitem{Andrews:2013uca} 
  M.~Andrews, K.~Hinterbichler, J.~Stokes and M.~Trodden,
  Class.\ Quant.\ Grav.\  {\bf 30}, 184006 (2013)
  [arXiv:1306.5743 [hep-th]].

\bibitem{Goon:2014ywa} 
  G.~Goon, A.~E.~Gumrukcuoglu, K.~Hinterbichler, S.~Mukohyama and M.~Trodden,
  arXiv:1402.5424 [hep-th].

\bibitem{Rubakov:2004eb} 
  V.~A.~Rubakov,
  hep-th/0407104.

\bibitem{Dubovsky:2004sg} 
  S.~L.~Dubovsky,
  JHEP {\bf 0410}, 076 (2004)
  [hep-th/0409124].

\bibitem{Blas:2009my} 
  D.~Blas, D.~Comelli, F.~Nesti and L.~Pilo,
  Phys.\ Rev.\ D {\bf 80}, 044025 (2009)
  [arXiv:0905.1699 [hep-th]].

\bibitem{Comelli:2013paa} 
  D.~Comelli, F.~Nesti and L.~Pilo,
  Phys.\ Rev.\ D {\bf 87}, 124021 (2013)
  [arXiv:1302.4447 [hep-th]].

\bibitem{Comelli:2013txa}
  D.~Comelli, F.~Nesti and L.~Pilo,
  JHEP {\bf 1307}, 161 (2013)
  [arXiv:1305.0236 [hep-th]].

\bibitem{Comelli:2013tja}
  D.~Comelli, F.~Nesti and L.~Pilo,
  JCAP {\bf 1405}, 036 (2014)
  [arXiv:1307.8329 [hep-th]].

\bibitem{ArkaniHamed:2003uy} 
  N.~Arkani-Hamed, H.~-C.~Cheng, M.~A.~Luty and S.~Mukohyama,
  JHEP {\bf 0405}, 074 (2004)
  [hep-th/0312099].

\bibitem{deRham:2013qqa} 
  C.~de Rham, L.~Heisenberg and R.~H.~Ribeiro,
  Phys.\ Rev.\ D {\bf 88}, 084058 (2013)
  [arXiv:1307.7169 [hep-th]].

\end{thebibliography}
\end{document}